\documentclass[a4paper,fleqn]{cas-dc}

\usepackage[numbers]{natbib}
\usepackage{silence}
\usepackage{caption}
\usepackage{lastpage}
\def\tsc#1{\csdef{#1}{\textsc{\lowercase{#1}}\xspace}}
\tsc{WGM}
\tsc{QE}
\begin{document}
\def\lastpage{\pageref{LastPage}}
\let\WriteBookmarks\relax
\def\floatpagepagefraction{1}
\def\textpagefraction{.001}

\shorttitle{Gamma-ray millisecond pulsars in globular clusters}    

\shortauthors{Lawson \& Lorimer}  

\title [mode = title]{Gamma-ray and radio populations of millisecond pulsars in globular clusters}  

\tnotemark[1] 

\tnotetext[1]{This document is the results of the research project funded by the National Science Foundation.} 


\author[1]{Hannah Lawson}[orcid=0009-0004-9304-8325]%[<options>]

\cormark[1]

\fnmark[1]

\ead{hannahlawson@ucsb.edu}

\ead[url]{https://hcl2k04.wixsite.com/hannahlawson}

\credit{Data curation, Formal analysis, Investigation, Methodology, Software, Validation, Visualization, Writing – original draft, Writing – review and editing}


\author[2,3,4]{Duncan R. Lorimer}[orcid=0000-0003-1301-966X]%[]

% Footnote of the second author

% Email id of the second author
\ead{duncan.lorimer@mail.wvu.edu}

\ead[url]{https://physics.wvu.edu/directory/faculty/duncan-lorimer}

\credit{Conceptualization, Formal analysis, Funding acquisition, Methodology, Supervision, Validation, Writing – original draft, Writing – review and editing}

\affiliation[1]{organization={The University of Texas at Austin, Department of Astronomy},
            addressline={2515 Speedway}, 
            city={Austin},
%          citysep={}, % Uncomment if no comma needed between city and postcode
            postcode={78712}, 
            state={TX},
            country={USA}}

\affiliation[2]{organization={Department of Physics and Astronomy, West Virginia University},
            addressline={P.O. Box 6315}, 
            city={Morgantown},
%          citysep={}, % Uncomment if no comma needed between city and postcode
            postcode={26506}, 
            state={WVU},
            country={USA}}

\affiliation[3]{organization={Center for Gravitational Waves and Cosmology},
            addressline={Chestnut Ridge Building}, 
            city={Morgantown},
%          citysep={}, % Uncomment if no comma needed between city and postcode
            postcode={26506}, 
            state={WV},
            country={USA}}

\affiliation[4]{organization={National Centre for Radio Astrophysics, Tata Institute of Fundamental Research},
            addressline={Pune University Campus}, 
            city={Pune},
%          citysep={}, % Uncomment if no comma needed between city and postcode
            postcode={410 007}, 
            state={India},
            country={USA}}

\cortext[1]{Corresponding author}

\fntext[1]{Present address: Department of Physics, University of California, Santa Barbara, Santa Barbara, CA 93106, USA}


\begin{abstract}
The populations of millisecond pulsars (MSPs) across the Milky Way's globular cluster (GC) system serve as powerful diagnostics of neutron star evolution in a variety of dense stellar environments. Using properties of 118 Galactic GCs, we performed Monte Carlo simulations of their gamma-ray fluxes observed by {\it Fermi}. We compared two luminosity models: a 3D fundamental plane (model A) and a lognormal distribution (model B). Both models reproduce the observed gamma-ray fluxes and mirror the scaling law with the GC stellar encounter rate established for low-mass X-ray binaries and radio MSPs. Due to its lower average gamma-ray luminosities, model B predicts a 7--8 times larger MSP population ($5150^{+720}_{-710}$) than  model A ($670^{+70}_{-80}$). Furthermore, both models accurately predict a population-wide expectation of roughly four clusters whose aggregate gamma-ray emission is dominated by a single bright MSP. Invoking a radio-to-gamma-ray luminosity scaling ($L_r \sim 1 \times 10^{-6} L_{\gamma}$), we find that model B predicts a mean radio pseudoluminosity ($\log L_{\rm pseudo} \sim -1.1$) comparable to canonical pulsars.  In contrast, model A implies an intrinsically brighter population ($\log L_{\rm pseudo} \sim 0.2$) that lacks a clear physical basis and is inconsistent with the results found for Galactic pulsars. The larger underlying population of faint millisecond pulsars predicted by model B could be probed by sensitive surveys with emerging radio facilities to further constrain the evolutionary pathways of MSPs.
\end{abstract}


%\item Monte Carlo simulations of gamma-ray millisecond pulsars in globular clusters.
%\item Compared 3D fundamental plane (A) and lognormal (B) gamma-ray luminosity models.
%\item Model B predicts 5150 pulsars, 7--8 times larger than model A.
%\item For both models, a few cluster fluxes can be dominated by a single pulsar.
%\item Radio pulsar luminosity constraints favor model B over model A.

%\nocite{*}

% Keywords

\maketitle

% Main text
\section{Introduction}\label{introduction}

Following the discovery of the original millisecond pulsar \citep[MSP;][]{1982Natur.300..615B}, it has long been suspected that the majority, if not all, of these rapidly spinning objects are formed via mass transfer in low-mass X-ray binaries, where a neutron star is effectively `recycled' as accretion spins it up to millisecond periods \citep[for a review, see][]{2023pbse.book.....T}. Indeed, the formation of MSPs is known to be highly efficient in globular clusters (GCs), due to the high specific incidence of X-ray binaries in GCs \citep{1975ApJ...199L.143C}. Since the discovery of the first GC MSP \cite{1987Natur.328..399L}, thanks to substantial improvements in both telescope sensitivity and algorithmic development, the current population stands at 369 pulsars in 46 GCs\footnote[2]{\url{https://www3.mpifr-bonn.mpg.de/staff/pfreire/GCpsr.html}}. This sample, where over 90\% of all pulsars have periods less than 30~ms, predominantly consists of MSPs. Some of these are known to be gamma-ray emitters \citep[see, e.g.,][]{2023ApJ...958..191S}. While the handful of canonical (i.e.~non-recycled) pulsars in GCs is an interesting area of study \citep[see, e.g.,][]{2011ApJ...742...51B}, canonical pulsars are energetically subdominant gamma-ray emitters compared to the much larger sample of MSPs in GCs which is the focus of this paper.

Our current knowledge of the MSP population in general is extensive, thanks to a large number of previous studies both of the Galactic disk population as well as those MSPs across the Galactic system of GCs \citep[for recent reviews, see][]{2025OJAp....854653L,2025OJAp....854251B}. Most of this work has been carried out at radio wavelengths, where many large-scale and targeted MSP surveys have been conducted in the past 40 years. In the Galactic disk, previous studies have constrained the MSP period distribution \citep{2015MNRAS.450.2185L}, luminosity function \citep{2013MNRAS.433..259B}, scale height and Galactic distribution \citep{gonthier2018}. In GCs, often informed by prior constraints from Galactic MSPs, significant progress has been made in constraining the pulsar content (hereafter referred to as abundance) in each GC. From constraints on MSP abundances in 9 GCs, \citet{hui2010} found that the abundance of potentially observable radio MSPs scales with stellar encounter rate, $\Gamma$, roughly as $\Gamma^{2/3}$ with a similar dependence on GC metallicity. The dependence of abundance on $\Gamma$ was confirmed by \citet{turk2013} in a complementary study to include a sample of MSPs across 94 GCs. This analysis assumed the underlying radio luminosity function to be lognormal in form, as found by \citet{2006ApJ...643..332F} for Galactic pulsars and supported by \citet{2011MNRAS.418..477B} in their study of GC MSPs. \citet{2013MNRAS.431..874C} subsequently developed a Bayesian framework to constrain the radio abundance and luminosity function for GCs with significant numbers of observed MSPs and measurements of the diffuse radio flux. However, as discussed by \citet{2024ApJ...974..144B} in their implementation of this approach, current samples of GCs are not probed deeply enough to robustly constrain the MSP population over a large number of GCs. We note that deep continuum imaging of individual well-populated clusters such as Terzan~5 suggests MSP populations substantially larger than previously inferred \citep{2026MNRAS.547ag276U}.

While the radio population provides a foundational understanding of pulsar abundances, recent multi-wavelength studies have begun to offer a more holistic view of MSP energetics. Notably, \citet{2023ApJ...944..225L} conducted a comparative analysis of Galactic field and GC MSPs, finding that while the X-ray luminosity functions of GC and Galactic MSPs are broadly similar, the GC population exhibits a higher proportion of so-called `spider' binaries \citep[for a review of these systems see, e.g.,][]{2013IAUS..291..127R} which significantly impact the observed X-ray emission. This perspective is further bolstered by observations in the gamma-ray regime. The cumulative contribution of pulsars to the gamma-ray sky has long been a subject of interest, dating back to early work of \citet{1992ApJ...391..659B} regarding the pulsar contribution to the Galactic gamma-ray background. Following the launch of the {\it Fermi} Gamma-ray Space Telescope, \citet{Abdo2010} presented the first systematic population study of gamma-ray emission from Galactic GCs, establishing the initial empirical link between a GC's aggregate gamma-ray luminosity and its stellar encounter rate. More recently, \citet{gonthier2018} utilized {\it Fermi} data to model the gamma-ray luminosity of both resolved and unresolved MSPs, demonstrating that their collective emission can account for the bulk of the observed gamma-ray flux from GCs. These studies suggest that gamma-ray observations may provide a more direct probe of the total MSP content in GCs than analyses of radio surveys alone.

In this paper, we describe and employ Monte Carlo simulations to estimate the number of gamma-ray MSPs in a sample of GCs. The framework of the simulations relies on using a gamma-ray MSP luminosity function to generate mock MSP fluxes, so we present two different models, each based on a different luminosity function from the existing literature. We present our sample of GCs and the input parameters required by our models in Section \ref{sec:sample}. We  describe the simulation procedure involving the two gamma-ray luminosity models in Section \ref{sec:mcsimulation}. In Section \ref{sec:results}, we present and discuss the results. We describe how the results scale with underlying GC properties, the differences between the results of the two gamma-ray luminosity models, and the implications of each model on the radio MSP content of the GCs in our sample that are well-detected in the radio. We present our main conclusions in Section \ref{sec:conc}.

% Use the style of numbering in square brackets.
% If nothing is used, default style will be taken.
%\begin{enumerate}[a)]
%\item 
%\item 
%\item 
%\end{enumerate}  

%\begin{itemize}
%\item 
%\item 
%\item 
%\end{itemize}  

%\begin{description}
%\item[]
%\item[] 
%\item[] 
%\end{description}  

% Uncomment and use as the case may be
%\begin{theorem} 
%\end{theorem}

% Uncomment and use as the case may be
%\begin{lemma} 
%\end{lemma}


\section{GC and MSP samples}\label{sec:sample}

We model a sample of 118 Galactic GCs for our analysis, using the parameters assembled in Table \ref{tab:Table_1}. As discussed in detail in Section \ref{sec:results}, we utilize the stellar encounter rates in our analyses, and the 118 GCs we tabulate are those for which a reliable stellar encounter rate, $\Gamma$, exist and we use the compilation of \citet{bahramian2013}. For consistency, we take the mean distances determined by \citet{baumgardt2021} for all 118 GCs in our sample as the distances to each GC, $D$, shown in this table. Since we expect GCs with higher $\Gamma$ values to have larger MSP abundances, we list the GCs in Table \ref{tab:Table_1} in descending order of $\Gamma$.

To estimate the MSP abundances in each GC, our model requires gamma-ray flux ($F_\gamma$) measurements or upper limits for each GC. For 37 GCs in our sample, we list flux measurements from the Large Area Telescope (LAT), the primary instrument on board the {\it Fermi} Gamma-Ray Space Telescope. For these GCs we extracted $F_\gamma$ values directly from the fourth data release\footnote[3]{\url{https://fermi.gsfc.nasa.gov/ssc/data/access/lat/14yr_catalog}} of the {\it Fermi} LAT source catalogue \citep[hereafter 4FGL-DR4;][]{2023arXiv230712546B} which includes 14 years of data for over 7,000 sources observed in the gamma-ray regime at energies\footnote[4]{We use eV throughout this work when referring to gamma-ray energies (1~eV is approximately equal to $1.6 \times 10^{-19}$~J).} ranging from 50 MeV to 1 TeV.

For the 81 GCs which were not detected by {\it Fermi} LAT, we obtain flux upper limits, as depicted by the $<$ symbols in Table \ref{tab:Table_1}. For GCs relying on upper limit measurements, the results quoted later will be an upper bound on the GC's MSP content. For {\it Fermi} LAT, which was used to generate the Third Pulsar Catalog \citep[3PC;][]{2023ApJ...958..191S}, an all-sky map of the flux sensitivity of this instrument is publicly available\footnote[5]{\url{https://fermi.gsfc.nasa.gov/ssc/data/access/lat/3rd_PSR_catalog}}. Using this map and the Galactic coordinates ($l$ and $b$) of each GC (obtained from the Sesame\footnote[6]{\url{https://cds.unistra.fr/cgi-bin/Sesame}} name resolver service), we find the LAT sensitivity at each GC's location. This was possible for 63 of the remaining GCs, so we set these sensitivities as a flux upper limit since they should have been detected by {\it Fermi} LAT had their fluxes been any greater. 

For the last 18 GCs, the {\it Fermi} LAT sensitivities at their locations were initially unknown. For the majority of these, we used the nearest object in 4FGL-DR4 within a 1-degree separation, taking the upper bound of its flux as a conservative upper limit. These flux upper limits are denoted by an asterisk in Table~\ref{tab:Table_1}, along with the name of the nearby 4FGL object. We note that these upper limits should be regarded with caution, as factors like background contamination near the Galactic plane or the presence of multiple nearby sources can complicate their reliability. However, for NGC 6624, NGC 6626, and NGC 6652, treating them as non-detections with unconstraining upper limits is incorrect. All three of these GCs host exceptionally high spin-down luminosity MSPs with firmly detected gamma-ray pulsations (PSR~B1821$-$24 in NGC~6626, \citealt{Wu2013}; PSR~J1823$-$3021A in NGC~6624, \citealt{Freire2011}; and PSR~J1835$-$3259B in NGC~6652, e.g., \citealt{Gautam2022, Zhang2022}). For these GCs, we instead explicitly adopt the measured flux of the bright millisecond pulsar as the aggregate GC flux and discuss how common such situations arise in the context of our modeling in Section \ref{sec:p1}. For the remaining GCs, setting the upper limit to be the flux of the nearest source is the most conservative approach in the absence of any further constraints.

To date, 45 of these GCs have known radio pulsar detections. For these, we note the number of radio pulsars $N_{\rm PSR,R}$ that have been observed in each GC. Some of the detected pulsars even have flux density measurements, either directly measured in or scaled to 1400~MHz, or the $S_{1400}$ band. For GCs with observed radio pulsars in which at least one flux measurement is known, we note the minimum detectable flux density detected so far to be the minimum $S_{1400}$ flux density, which we henceforth refer to as $S_{\rm min}$. The tabulated $S_{\rm min}$ values were obtained from the individual GC pulsar survey papers and kindly provided to us by K.~Halley. The tabulated $N_{\rm PSR,R}$ values were obtained from the online catalogue maintained by P.~Freire.

\section{Monte Carlo simulations}\label{sec:mcsimulation}

To constrain the number of gamma-ray MSPs in a given GC, we use  Monte Carlo methods to model the gamma-ray fluxes expected from a population of MSPs and confront these with measurements or upper limits on the fluxes from {\it Fermi} LAT. As discussed in Section \ref{sec:sample}, for those GCs that were not detected by {\it Fermi} LAT, we use either the instrumental sensitivity at the GC's Galactic coordinates or the flux of the closest 4FGL-DR4 object as the flux upper limit. For each of the 118 GCs in our sample, we simulate MSPs one at a time and compute their gamma-ray luminosities as described in detail below. The gamma-ray flux of the $i^{\rm th}$ MSP with luminosity $L_{\gamma,i}$, in GC $j$ is given by
\begin{equation}
\label{equ:fgamma}
    F_{\gamma,i} = \frac{L_{\gamma,i}}{4 \pi f_\gamma D_j^2},
\end{equation}
where $f_\gamma$ is the gamma-ray beaming fraction discussed below, and $D_j$ is the distance to the GC. We run each simulation until the sum of the MSP fluxes reaches the {\it Fermi} flux measurement or upper limit, counting up the number of MSPs that were required to do so. For each GC, we repeat this entire process 1000 times to obtain a distribution of the total number of MSPs in each GC. To account for the uncertainties in both the gamma-ray flux measured by {\it Fermi} LAT and the measured distance to each GC, for each iteration we vary the calculated fluxes and distances slightly by drawing them from a normal distribution with their errors as the standard deviations. However, we do not have uncertainties on the {\it Fermi} LAT sensitivities, so to maintain consistency, we do not vary any of the flux upper limits. After 1000 iterations, we report the median number of MSPs required for each GC as well as the 90\% confidence interval of the resulting population distribution.

Although the gamma-ray beaming fraction $f_{\gamma}$ for GC MSPs is uncertain, our use of isotropic-equivalent luminosity models renders our results independent of this choice. A beaming fraction $f_\gamma < 1$ increases the inferred intrinsic luminosity of individual pulsars, meaning proportionately fewer MSPs are required to match the observed GC flux. Consequently, when scaling the population back to account for beaming, the total inferred number of MSPs in each GC remains unchanged. We therefore set $f_\gamma = 1$ throughout this work, returning to beaming effects when simulating the radio population in Section \ref{sec:radio_model}.

\subsection{Model A: 3D fundamental plane} \label{sec:modela}

In their studies of the gamma-ray pulsar population, \citet{kalap2019,kalap2022} found the gamma-ray luminosity ($L_{\rm \gamma}$) to be strongly correlated with its dipole magnetic field strength ($B$), spin-down luminosity ($\dot E$), and spectral cutoff energy ($\epsilon_{\rm cut}$). Forming a plane in log-space, this relationship is known as three dimensional fundamental plane (3DFP). We adopt the latest version of the 3DFP using pulsars discovered by the 4FGL-DR3 in \citet{kalap2022}, henceforth referred to as model A, where
\begin{equation} \label{eq:3DFP}
\begin{split}    
        L_{\gamma} = 2.2 \times 10^{30} \, {\rm W} \, \bigg(\frac{B}{10^{4}~{\rm T}}\bigg)^{0.12} 
        \, \bigg(\frac{\dot{E}}{10^{24}~{\rm W}}\bigg)^{0.39} \\
        \bigg(\frac{\epsilon_{\rm cut}}{\rm MeV}\bigg)^{1.39}.
\end{split}
\end{equation}
To compute each of the parameters on the right-hand side of this equation, we first need to model the spin-down of each MSP. Following \citet{shawaiz2025}, this was done by drawing the age of each MSP ($t$) from a uniform distribution spanning 0--5~Gyr, the initial period ($P_{0}$) as a lognormal distribution with mean of 0.98~ms and standard deviation of 0.52~ms, and  birth magnetic field $B_{0}$ (T) with $\log B_0$  drawn from a normal distribution with a mean of 4.2 and standard deviation of 0.3. Using the spin-down model of \citet{2014ApJ...793...97K}, the period at time $t$,
\begin{equation}
\label{eq:P(t)}
%    P = \sqrt{P_{0}^2+\frac{2\pi^{2}R^{6}B_{0}^{2}}{c^3I}(1 + \sin^2\alpha)t},  CGS version
    P = \sqrt{P_0^2 + \frac{4 \pi \mu_0 R^6 B_{0}^2}{3 c^3 I} (1 + \sin^2\alpha)t},
\end{equation}
where $\mu_0$ is the permeability of free space, $R = 12$~km is the assumed neutron star radius, $c$ is the speed of light
and $I = 1.7 \times 10^{38}$~kg~m$^2$ is the moment of inertia and $\alpha$ is the magnetic inclination angle which we assume to be randomly oriented, i.e.~we draw from a distribution that is uniform in $\cos \alpha$. Differentiating Eq.~\ref{eq:P(t)}, the present day period derivative
\begin{equation}
    \dot{P} = \frac{P^2 -P_0^2}{2Pt}.
\end{equation}
Finally, for consistency with the implementation of Eq.~\ref{eq:3DFP} in our simulations, we use these $P$ and $\dot P$ values to compute  $B$ and $\dot E$, using the standard expressions \citep[]{lorimer2004}
given by
\begin{equation}
    B = 3.2 \times 10^{15} \, {\rm T}\, \left(\frac{P}{\rm s}\right)^{1/2}\, \left(\frac{\dot P}{\rm s/s}\right)^{1/2}
\end{equation}
and
\begin{equation}
    \dot E = 4.0 \times 10^{24} \, {\rm W}\, \bigg(\frac{P}{\rm s}\bigg)^{-3}\, \bigg(\frac{\dot P}{\rm s/s}\bigg).
\end{equation}
Having obtained $B$ and $\dot{E}$, following \cite{shawaiz2025} we draw the logarithm of the cut-off energy in log-space, sampling log$_{10} \epsilon_{\rm cut}$ from a normal distribution with a mean of 3.458 and standard deviation of 0.181. These parameters were obtained by fitting to the observed $\epsilon_{\rm cut}$ values for MSPs in the 4FGL-DR3 catalog. Finally, we compute the gamma-ray luminosity of the MSP with Eqn.~\ref{eq:3DFP}. 
To account for the scatter around the fundamental plane, we dither $L_\gamma$ by adding a random deviate drawn from a normal distribution with zero mean and standard deviation of 0.33~dex to the value of $\log_{10} L_\gamma$ obtained from Eqn.~\ref{eq:3DFP}. This value is motivated by the standard deviation of residuals around the best-fit plane reported by \citet{kalap2022} for their sample of 190 4FGL-DR3 MSPs.

\subsection{Model B: Lognormal} \label{sec:modelb}

In a more recent study, \citet{holst2025} used MSP detections from 3PC to model the gamma-ray luminosity of MSPs, proposing
that their luminosity function follows a lognormal distribution, i.e. the differential number of MSPs
\begin{equation} \label{eq:lognorm}
{{\rm d}N} \propto \frac{1}{L_{\gamma}} \exp \left (- \frac{(\ln L_{\gamma}-\ln L_{0})^2}{2 \sigma_{L}^2} \right)\, {\rm d}L_{\gamma},
\end{equation}
where  $L_{0}$, and $\sigma_{L}$ are free parameters characterizing the distribution which has a mean value
\begin{equation} \label{eq:lognorm_param}
\langle L_{\gamma} \rangle = L_{0} \exp \left ( \frac{\sigma_{L}^2}{2} \right ).
\end{equation}
\citet{holst2025} find $\langle L_{\gamma} \rangle = 5.9 \times 10^{25}$~W and $\sigma_{L}=2.3$. The best-fit mean of the distribution is, therefore, $\ln L_{0} = 56.7$. To implement this model (hereafter, model B) in our work, we sample the gamma-ray luminosity from a lognormal distribution with the aforementioned parameters and compute the corresponding contribution of each model MSP to the gamma-ray flux of its GC using Eqn.~\ref{equ:fgamma}.

\section{Results and Discussion} \label{sec:results}

\begin{figure}
    \centering
    \includegraphics[width=\linewidth]{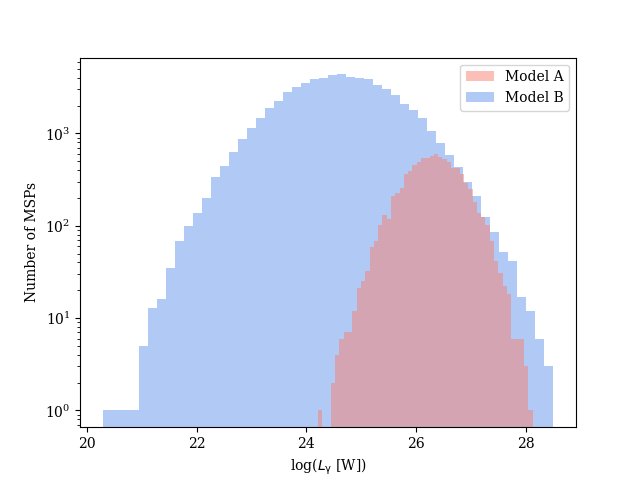}
    \caption{A comparison of the gamma-ray luminosities predicted by model A and B.}
    \label{fig:Figure_1}
\end{figure}

In the sections below, we present the results and discuss the differences between the two gamma-ray luminosity models: the 3DFP \citep[model A;][]{kalap2022} and the lognormal distribution \citep[model B;][]{holst2025}. The differences in luminosity between the two models are shown in Fig.~\ref{fig:Figure_1}.
As can be seen, model B has a broader range which expands that seen in model A to include potentially brighter and fainter MSPs. 
We discuss the consequences of these differences in the sections below.

\subsection{Population estimates from {\it Fermi} gamma-ray fluxes}\label{sec:npdfs}

\begin{figure*}
    \centering
    \includegraphics[width=\linewidth]{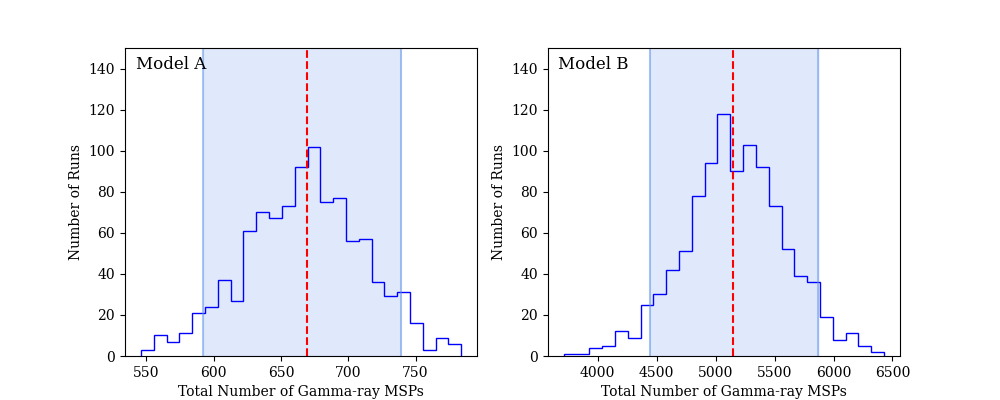}
    \caption{Simulation results showing the total number of GC MSPs predicted across the population of 118 GCs for model A (left) and model B (right). The dashed lines show the medians of each distribution, while the 90\% confidence intervals are marked as shaded regions: $670^{+70}_{-80}$ (model A) and $5150^{+720}_{-710}$ (model B).
    }
    \label{fig:Figure_2}
\end{figure*}

In Table \ref{tab:Table_2}, the median estimated number of MSPs that satisfies the gamma-ray flux or upper limit flux of each GC is shown for models A and B, respectively, as $N_{\gamma}^{A}$ and $N_{\gamma}^{B}$, along with the 90\% confidence intervals determined from the Monte Carlo runs. We use the $\gamma$ subscript here to indicate that these numbers have been obtained directly from simulations designed to mimic the gamma-ray flux measurement/limit from {\it Fermi} LAT as detailed above. While both models have been tuned to provide a good match to the observed gamma-ray flux of each GC, as shown in Table \ref{tab:Table_2}, the total number of gamma-ray pulsars across all GCs is fundamentally different: model B generally predicts \textbf{7--8} times more MSPs compared to model A. This result reflects the fact that the mean gamma-ray luminosity for the pulsars we generated in model A is typically \textbf{7--8} times larger than the corresponding sample in model B (Fig.~\ref{fig:Figure_1}). We see the impact of this in the total population of GC pulsars predicted by the two models in Figure \ref{fig:Figure_2}. For both models, the simulations are run 1000 times to show the distribution in the total number of MSPs predicted across all 118 GCs. Across these GCs as a whole, we find a total population of $670^{+70}_{-80}$ MSPs for model A and $5150^{+720}_{-710}$ MSPs for model B. It should be stressed at this point that both models account for the gamma-ray fluxes across a wide range of GC properties. We attempt to discriminate between the models later when we confront their predictions for radio MSPs.

\subsection{Abundance models and predicted gamma-ray fluxes} \label{sec:GC_abundance}

\begin{figure*}
    \centering
    \includegraphics[width=\linewidth]{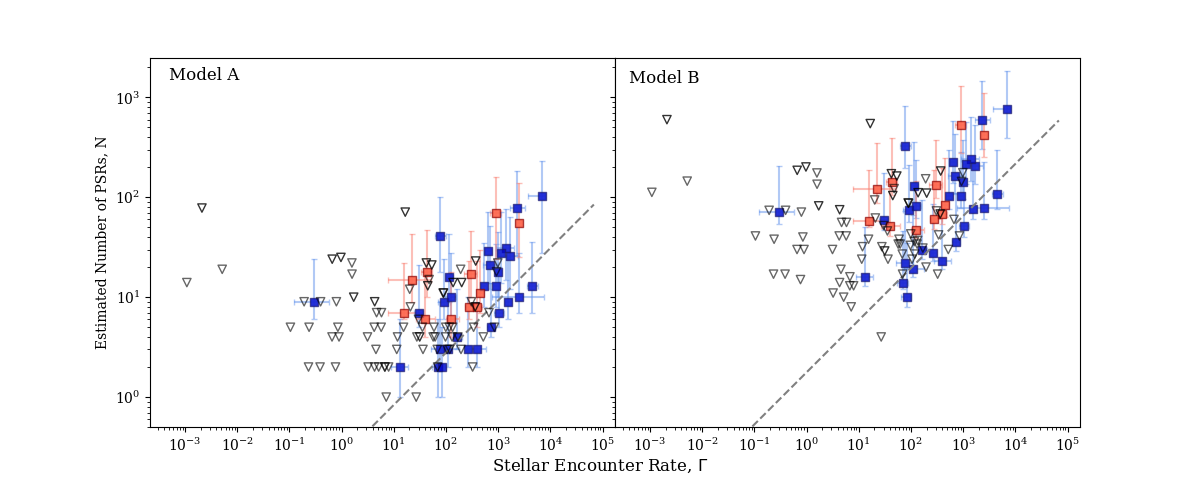}
    \caption{Fits for model A (left) and model B (right) to the scaling relationship $N = \alpha\Gamma^{\beta}$ obtained via a  MCMC that accounts for both detections (squares) and upper limits (triangles). GCs with no currently detected radio pulsars are colored red. The best-fit relations are $N = 0.25^{+0.11}_{-0.08}\,\Gamma^{0.52^{+0.06}_{-0.06}}$ for model A and $N = 1.77^{+0.93}_{-0.67}\,\Gamma^{0.52^{+0.08}_{-0.07}}$ for model B, respectively.} 
    \label{fig:Figure_3}
\end{figure*}

\citet{hui2010} and \citet{turk2013} found that the MSP abundance in a GC, $N$, scales with the stellar encounter rate, $\Gamma$, as $N \propto \Gamma ^{2/3}$. As shown in Fig.~\ref{fig:Figure_3}, we also find that the number of pulsars predicted by both models scales in this way for the MSP abundances predicted by both model A and model B. To better understand the implications of our two luminosity models, we fit GC abundance models of the form $N = \alpha \Gamma^{\beta}$ to the $N_\gamma$ values, treating detections and upper limits using a Markov Chain Monte Carlo (MCMC) approach. Each GC with a {\it Fermi} LAT detection contributes a Gaussian log-likelihood term centred on its median $N_\gamma$ with width derived from the 90\% confidence interval, while each upper-limit GC contributes a term encoding the probability that the true $N$ lies below the 95th percentile of its upper-limit distribution. We allow the parameters $\alpha$ and $\beta$ to vary freely. For both models, the MCMC posteriors yield $\beta = 0.52$, with best-fit normalizations $\alpha_A = 0.25^{+0.11}_{-0.08}$ and $\alpha_B = 1.77^{+0.93}_{-0.67}$. The fitted exponent of 0.52 is shallower than the value of $2/3 \approx 0.67$ found by \citet{hui2010} and \citet{turk2013}, though $2/3$ lies approximately $2\sigma$ from the posterior median and cannot be firmly excluded with the present data. Notably, both models differ only in normalization by a factor of $1.77/0.25 \approx 7$, consistent with the factor of \textbf{7--8} difference in total predicted MSP population discussed above.

Apart from the relatively unconstrained upper limits shown in Figure~\ref{fig:Figure_3}, a particularly notable outlier is Terzan~1, which has a very low stellar encounter rate of approximately 0.3 but features a high predicted pulsar abundance. It is possible that the stellar encounter rate for Terzan~1 is significantly underestimated due to heavy interstellar extinction and severe stellar crowding given its close projection to the Galactic Center. Alternatively, as discussed by \citet{2006MNRAS.369..407C}, Terzan~1 may be a core-collapsed GC that has lost a substantial fraction of its stellar mass via tidal stripping while retaining its central binary and pulsar populations. Terzan~1 is currently known to host six pulsars with spin periods less than 20~ms \citep{2024arXiv241211271S}.

\begin{figure*}
    \centering
    \includegraphics[width=\linewidth]{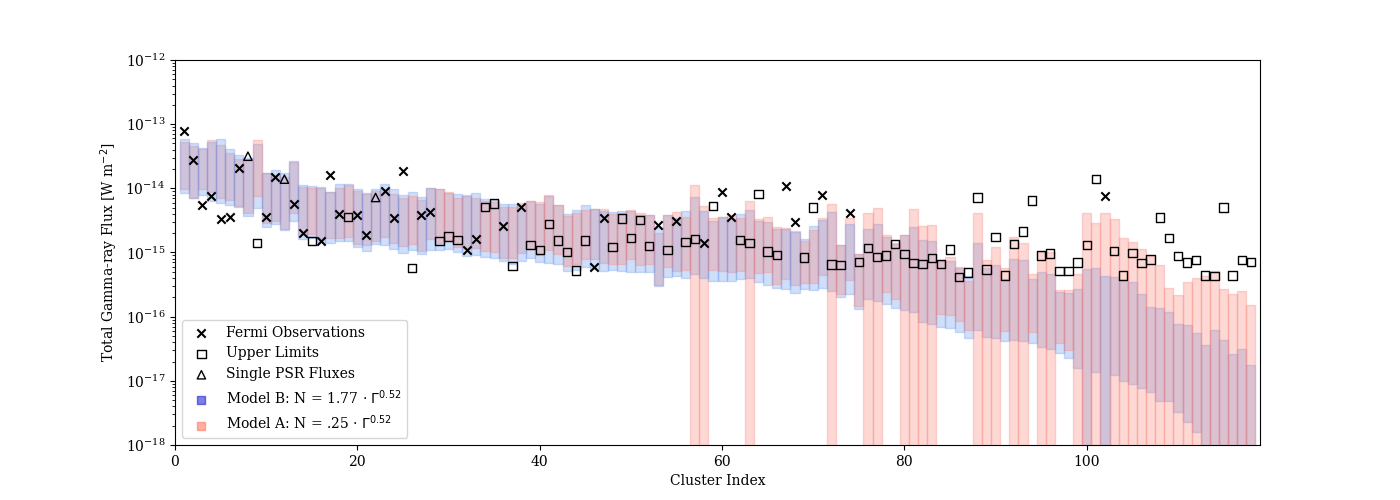}
    \caption{Predicted gamma-ray fluxes for each GC based on the scaling laws we derived in Section \ref{sec:GC_abundance} for model A (red) and model B (blue), with the 90\% confidence intervals shaded. We also show the {\it Fermi} LAT measured GC fluxes (black crosses), measured bright MSP fluxes (black triangles), and upper limits (black squares) to demonstrate how well the modeled fluxes agree with these values. The GCs appear in descending order according to the median flux predicted by model B.}
    \label{fig:Figure_4}
\end{figure*}

To demonstrate the utility of these scaling laws, and as a self-consistency check, we apply the aforementioned fits from both
models A and B to {\it predict} the gamma-ray flux for each GC. This is essentially the same process as described in Section \ref{sec:mcsimulation}, but we determine how many MSPs to generate by drawing $N$ from a normal distribution, with the mean as predicted by the abundance model and the error propagated through the model as the standard deviation. 
The results of 
this process for both models A and B are shown in Fig.~\ref{fig:Figure_4}. For model A, 92 out of 118 GCs have predicted fluxes that are in agreement with the measured fluxes and estimated upper limits. For model B, 97 GCs are in agreement with these values. For the observed fluxes, agreement means that the error bars of the observed {\it Fermi} LAT flux (or the measured MSP flux for NGC 6626, NGC 6624, and NGC 6652) at least overlap with the error bars of the flux predicted by the model. For the upper limit fluxes, agreement is defined as when the lower error of the predicted flux is indeed smaller than the upper limit estimation (either from the 3PC sensitivity or 4FGL object flux), meaning that we would not have expected to be able to observe this GC.

\begin{figure*}
    \centering
    \includegraphics[width=0.48\linewidth]{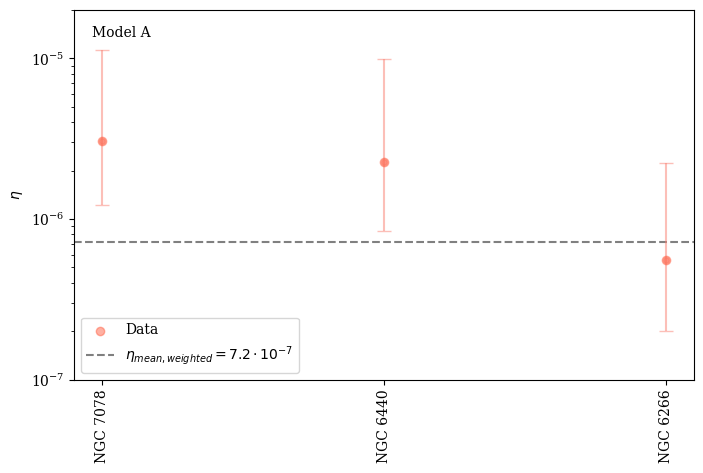}
    \hfill
    \includegraphics[width=0.48\linewidth]{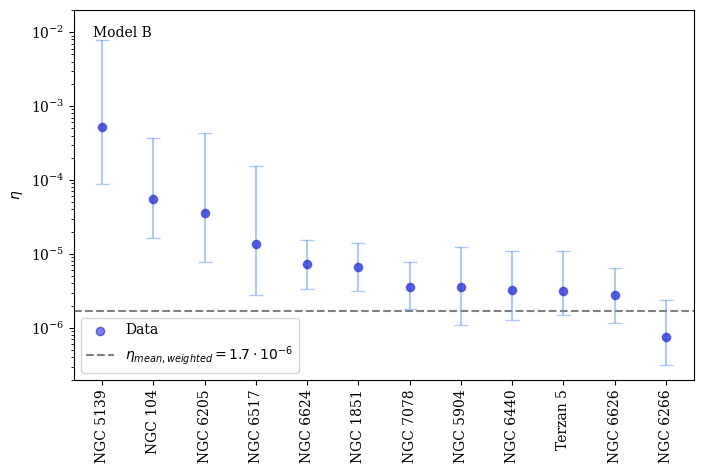}
    \caption{Results of simulations to determine the relative radio-to-gamma-ray luminosity ratio, $\eta$ for model A (left) and model B (right). For \textbf{eight} of the GCs that had at least five known radio pulsars with measured radio flux densities, we were unable to estimate $\eta$ for model A because this model estimated a total number of pulsars that was less than the number of known pulsars for that GC. Model B predicted a sufficiently large population to estimate $\eta$ in all cases.} 
    \label{fig:Figure_5}
\end{figure*}

\subsection{GCs dominated by a single bright MSP}\label{sec:p1}

As mentioned in Section \ref{sec:sample}, for at least three GCs (NGC~6624, NGC~6652 and NGC~6626) the tabulated gamma-ray flux is dominated by a single MSP. To estimate how likely this is to occur in the population as a whole, we computed the fraction of simulation runs across 1000 realizations where a GC's brightest individual pulsar exceeds the observed total flux ($p_1^A$ for model A and $p_1^B$ for model B, tabulated in Table~\ref{tab:Table_2}). 
The $p_1$ values span a wide range across our sample, from well below 1\% for richly populated GCs to above 50\% for GCs with very few predicted MSPs (e.g., NGC~6397, NGC~6205, and NGC~6752). While model B consistently predicts higher fractions than model A due to its broader luminosity function and higher draw frequency, both models demonstrate that single-pulsar dominance is statistically expected overall. The expected total number of GCs exhibiting single-pulsar dominance is $\langle N_{\rm dom}\rangle = \sum_j p_{1,j}$, where the sum runs over all GCs with known radio pulsars. Evaluating this sum yields roughly 4 such cases for either model, consistent with the small number of currently observed cases. That we do not predict the specific individual GCs where this occurs is a natural consequence of the simulations’ stochastic nature: the models provide statistical expectations across probability space, whereas the actual Milky Way GC system represents only a single realized outcome.

\begin{figure}
    \centering
    \includegraphics[width=\linewidth]{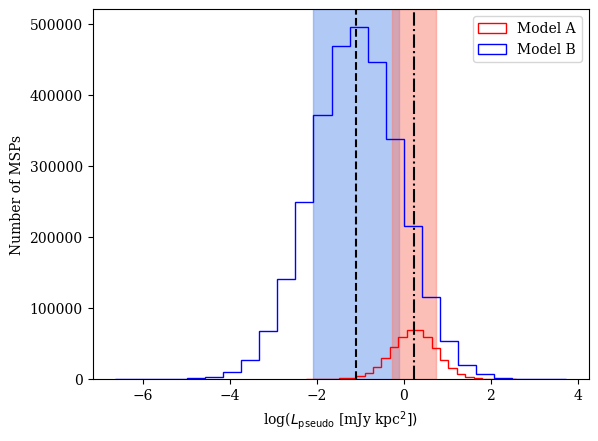}
    \caption{Pseudoluminosities for the radio MSP populations predicted for models A (red) and B (blue). The vertical lines show the mean
    ${\rm log}$ $L_{\rm pseudo}$ for each model. The shaded regions show the 1$\sigma$ confidence intervals of the two distributions. The respective means and standard deviations in $\log L_{\rm pseudo}$ of 0.2 and 0.5 for model A and --1.1 and 1.0 for model B.}
    \label{fig:Figure_6}
\end{figure}

\subsection{Total GC pulsar populations and the number of observable radio pulsars}\label{sec:radio_model}

Thus far we have seen that both models A and B can be used with their respective GC abundance models to reasonably predict
the observed gamma-ray fluxes seen with {\it Fermi} LAT across the GC population. Fundamentally, however, each gamma-ray luminosity model has very different implications for the total population of gamma-ray MSPs across all GCs in our sample. Due to its smaller average gamma-ray luminosity, model B requires a total MSP population that is around \textbf{7--8} times that of model A.

In an effort to distinguish between these two models, we used them to jointly predict the number of gamma-ray and radio MSPs beaming towards the Earth. To do this, motivated by radio studies in which it is found that the radio luminosity, $L_r$, ranges between $10^{-5}$ and $10^{-8}$ of $\dot{E}$ \citep{2014ApJ...784...59S}, we adopt a simple model where
\begin{equation}\label{eq:radiogamma}
   L_r = \eta L_{\gamma}.
\end{equation}
Here, $\eta$ is a dimensionless scaling factor between radio and gamma-ray power output which we might expect to be in the range $10^{-4}$ to $10^{-7}$. To determine viable values for $\eta$ for each of model A and B, using only GCs with at least five known radio pulsars with measured radio flux densities, we draw populations of gamma-ray pulsars for each GC using the scaling relationship  
found above. Using these, we calculate the radio flux of each model pulsar
at 1400~MHz, $S_{1400}$. In this framework,
\begin{equation} \label{eq:radioflux}
S_{1400} = \frac{ \eta L_{\gamma}}{4 \pi f_r D^2 \Delta \nu},
\end{equation}
where $L_{\gamma}$ is the gamma-ray luminosity from either model A or B, $f_r$ is the radio beaming fraction, $D$ is the distance to the GC, and $\Delta \nu$ is the width of radio frequency
band. As can be seen in the above expression, in the lack of any other information, $f_r$ and $\eta$ are covariant. Guided by the results of previous work \citep{kramer1998}, we choose to set $f_r=0.75$. Finally, adopting $\Delta \nu = 800$~MHz (similar to that used in GC searches using the Green Bank Telescope at this frequency), we find
\begin{equation} \label{eq:radioflux2}
S_{1400} \simeq 1.75\, {\rm mJy} \, 
\bigg(\frac{L_\gamma}{10^{32}\,{\rm W}}\bigg) \;
\bigg(\frac{\eta}{10^{-6}}\bigg) \;
\bigg(\frac{D}{\rm kpc}\bigg)^{-2},
\end{equation}
where 1~Jy~$\equiv 10^{-26}$~W~m$^{-2}$~Hz$^{-1}$.
We initially set $\eta=1$ and na\"{\i}vely compute the $S_{\rm 1400}$ values of our model radio pulsars in
each GC. To account for the radio MSP beaming model of \cite{kramer1998}, we randomly reject 25\% of these model pulsars. In each 
realization for a given GC, we then rescale the radio fluxes by an appropriate value of $\eta$ such that
the number of radio pulsars above the minimum detectable radio threshold, $S_{\rm min}$, matches the observed number of
detections for that GC. For simplicity, for each GC, we choose $S_{\rm min}$ to be the flux of the faintest observable radio MSP.
This process is repeated 1000 times to get a distribution of $\eta$ for
each GC. These values are shown in Fig.~\ref{fig:Figure_5}. While there is some variation in 
$\eta$ values across GCs, the weighted mean values are $\eta_A=7.2 \times 10^{-7}$ for model A
and $\eta_B=1.7 \times 10^{-6}$ for model B. We note that for model~A, the estimated MSP population fell below the number of known radio pulsars for all but three of the GCs used in this analysis, meaning $\eta_A$ is derived from a very limited sample and should be interpreted with caution.

Assuming these weighted mean values across all GCs, we predict
the number of observable radio MSPs whose fluxes are greater than the minimum $S_{1400}$ currently observed
in each GC. These values are given in Table \ref{tab:Table_2} as $N^{A}_{\rm Radio}$ and $N^{B}_{\rm Radio}$ for
models A and B, respectively. As expected, both of these predictions generally track the current number of
radio pulsars observed in each GC. Deviations in individual GCs (see, e.g., the entry for NGC~104, aka 47~Tucanae)
can be attributed to our assumption of a single $\eta$ value for all GCs as well as details of individual
surveys (e.g., scintillation boosting of radio fluxes) that we do not capture in this approach. For NGC~104, the number of observed radio pulsars currently stands at 42 \citep{2026A&A...710A.130C}. 
This is almost double the total number
predicted by model A. A similar result is seen for NGC~5139 (Omega Centauri) where model A cannot explain the currently observed
radio MSP sample \cite{2023MNRAS.520.3847C}. We note that for Terzan~5, our model~B prediction of $N_\gamma^B = 770^{+300}_{-390}$ is broadly consistent with the recent finding of \citet{2026MNRAS.547ag276U}, who infer a minimum population of $\sim$250 detectable MSPs in that cluster. In contrast, our model~A prediction of $N_\gamma^A = 100^{+27}_{-28}$ falls significantly below this observational lower bound, providing further independent evidence in favour of model~B.

\subsection{Implications for radio MSP luminosities}

So far, our attempts to distinguish between model A (the 3DFP model) or model B (the lognormal model) for the 
gamma-ray luminosity function of MSPs have been unsuccessful. As shown in Fig.~\ref{fig:Figure_1}, model A predicts
generally higher gamma-ray luminosities and requires fewer MSPs in total than model B. We have been able to 
broadly account for the number of radio pulsars by invoking a simple radio-to-gamma luminosity model
given in Eq.~\ref{eq:radiogamma} with an efficiency parameter \textbf{$\eta \sim 10^{-6}$}. 

A very important consequence
of this approach is that the {\it underlying} distributions of radio luminosities reflect the
fundamental differences in gamma-ray luminosity between the two models. This can be clearly seen in Fig.~\ref{fig:Figure_6}
where we show the underlying distributions of luminosities for GC MSPs across all GCs, as
predicted by our models. In addition to showing the different sizes of each population, this figure shows clear differences
in the range of luminosities predicted by each model.
Here, following \citet{1997ApJ...482..971C}, we specifically use the radio `pseudoluminosity' $L_{\rm pseudo}=S_{1400}D^2$,
a commonly-used metric in radio population studies \citep[see, e.g.,][]{2006ApJ...643..332F}. As shown in Fig.~\ref{fig:Figure_6}, while
both distributions with $L_{\rm pseudo}>1$~mJy~kpc$^2$ are similar, the mean $L_{\rm pseudo}$ is significantly larger in model A than model B. While both models are therefore consistent with the small number of bright MSPs currently detected in GCs, the models diverge significantly at the low-luminosity end. 
Model B predicts a mean $\log L_{\rm pseudo} \sim -1.1$ with a standard deviation of $\sim$1.0~dex, consistent with both the mean and width of the canonical pulsar luminosity distribution found by \citet{2006ApJ...643..332F} ($\log L_{\rm pseudo} \sim -1.1$, $\sigma \sim 0.9$~dex). In contrast, model A predicts a mean $\log L_{\rm pseudo} \sim 0.2$, implying a population of GC MSPs that is more than an order of magnitude brighter than canonical pulsars.

The findings of \citet{2024MNRAS.532.3558K} provide important context for interpreting these results. They demonstrate that MSPs typically exhibit lower pseudoluminosities than canonical pulsars due purely to geometric effects, meaning the underlying MSP luminosity distribution should lie at or below the canonical pulsar mean. Model A’s prediction of a population more than an order of magnitude brighter than canonical pulsars seems clearly at odds with what we know about radio pulsar luminosities in general. Model B, by contrast, is broadly consistent with the radio luminosity distributions of both canonical and millisecond pulsars. We acknowledge that our treatment of the radio luminosities is intentionally simple, and more sophisticated population synthesis studies would be needed to make a fully rigorous comparison. Nevertheless, on the basis of the available evidence it seems very hard to reconcile model A with the observed properties of the radio pulsar population, and we conclude that model B provides a more physically motivated description of the underlying GC MSP population.

\section{Conclusions} \label{sec:conc}

We have performed an analysis of two different gamma-ray luminosity models to describe the underlying population of millisecond pulsars (MSPs) in the Galactic globular cluster (GC) system. Our primary conclusions are as follows:
\begin{itemize}
    \item While both the 3D fundamental plane (model A) and the lognormal luminosity model (model B) are capable of reproducing the gamma-ray flux observed by {\it Fermi} LAT, they predict fundamentally different numbers of MSPs: $670^{+70}_{-80}$ for model A and $5150^{+720}_{-710}$ for model B. Model B requires a \textbf{7--8} times larger population than model A due to its lower average gamma-ray luminosity.
    
    \item We demonstrate that the estimated number of pulsars in each GC scales with the stellar encounter rate as approximately $\Gamma^{1/2}$, somewhat shallower than but consistent within $\sim 2\sigma$ of the $\Gamma^{2/3}$ scaling found in earlier studies.

    \item Stochastic modeling of the GC population shows that aggregate gamma-ray emission can be dominated by a single bright MSP. Summing the fractional probabilities over all GCs with known radio pulsars yields an expected total of approximately 3 such systems for both models, aligning with current observations (NGC~6624, NGC~6652, and NGC~6626). The inability of the models to predict the exact GCs exhibiting this dominance is a natural consequence of the models reflecting statistical expectations rather than a specific deterministic outcome.
    
    \item Adopting a radio-to-gamma-ray luminosity scaling \textbf{($L_r  \sim 10^{-6} L_{\gamma}$)}, we find that both models are broadly consistent with the number of observed radio MSPs in GCs. Notable exceptions occur, however, for 47~Tucanae and Omega Centauri 
    where model A cannot account for the number of radio MSPs.
    
    \item The radio luminosity predictions for both models diverge significantly at the low-luminosity end. Model B predicts a pseudoluminosity that is consistent with both the mean and width of the canonical pulsar luminosity function \citep{2006ApJ...643..332F}. In contrast, model A requires a population more than an order of magnitude brighter and is very difficult to reconcile with what is known about radio pulsar luminosities. We conclude that model B provides a more physically motivated description of the underlying GC MSP population.
\end{itemize}

Future work will benefit from the increased sensitivity of next-generation facilities. Upcoming surveys of GCs with both the Deep Synoptic Array \citep{2026AAS...24831701H} and the Square Kilometre Array \citep{2025OJAp....854251B} and ongoing deep searches with MeerKAT and FAST \citep[see, e.g.,][]{2025SciBu..70.1568Z} will be essential to probe the low-luminosity regime where our models diverge. Constraining the population of faint MSPs would provide more definitive support for a significant population of lower luminosity MSPs, as suggested by model B. Incorporating more sophisticated geometric constraints such as varying magnetic inclination and observer angles into the Monte Carlo simulations could further refine the relationship between $L_{\rm pseudo}$ and intrinsic luminosity. Our treatment of radio selection effects was intentionally simplistic throughout this work. Applying improved population synthesis techniques to well-sampled GCs like Terzan~5 and 47~Tucanae would allow for a more detailed tests of the luminosity function parameters and the radio-to-gamma-ray efficiency. In this context, the recent deep continuum imaging of Terzan~5 by \citet{2026MNRAS.547ag276U}, which places a minimum population of $\sim$250 detectable pulsars in that cluster, already provides a strong independent constraint that favours model~B over model~A.

\section*{Acknowledgements}

This research was carried out during a 10-week Research Experience for Undergraduates program at West Virginia University. We acknowledge the National Science Foundation’s support of this program under award number 2348764. We thank Shawaiz Tabassum, Craig Heinke and David Smith for extremely useful suggestions and Abyss Halley for a compilation of radio pulsar flux density measurements.
We thank the referee for extremely useful comments on an earlier version of this manuscript.

\printcredits

%\bibliographystyle{cas-model1-num-names}
\bibliographystyle{cas-model2-names}

\bibliography{cas-refs}

@STRING{aap = "Astronom. Astrophys."}

@INPROCEEDINGS{2026AAS...24831701H,
       author = {{Hallinan}, Gregg},
        title = "{DSA: Key Science and Reference Design}",
    booktitle = {American Astronomical Society Meeting Abstracts},
         year = 2026,
       series = {American Astronomical Society Meeting Abstracts},
       volume = {58},
        month = jul,
          eid = {317.01},
        pages = {317.01},
       adsurl = {https://ui.adsabs.harvard.edu/abs/2026AAS...24831701H}
}

@ARTICLE{2026MNRAS.547ag276U,
       author = {{Urquhart}, Ryan and {Strader}, Jay and {Chomiuk}, Laura and {Ransom}, Scott M. and {Heinke}, Craig O. and {Bahramian}, Arash and {Maccarone}, Thomas J.},
        title = "{New deep radio continuum imaging still indicates a large reservoir of undiscovered millisecond pulsars in Terzan 5}",
      journal = {\mnras},
         year = 2026,
        month = apr,
       volume = {547},
       number = {2},
          eid = {stag276},
        pages = {stag276},
          doi = {10.1093/mnras/stag276},
archivePrefix = {arXiv},
       eprint = {2602.10199},
 primaryClass = {astro-ph.HE},
       adsurl = {https://ui.adsabs.harvard.edu/abs/2026MNRAS.547ag276U}
}

@ARTICLE{2023MNRAS.520.3847C,
       author = {{Chen}, W. and {Freire}, P.~C.~C. and {Ridolfi}, A. and {Barr}, E.~D. and {Stappers}, B. and {Kramer}, M. and {Possenti}, A. and {Ransom}, S.~M. and {Levin}, L. and {Breton}, R.~P. and {Burgay}, M. and {Camilo}, F. and {Buchner}, S. and {Champion}, D.~J. and {Abbate}, F. and {Venkatraman Krishnan}, V. and {Padmanabh}, P.~V. and {Gautam}, T. and {Vleeschower}, L. and {Geyer}, M. and {Grie{\ss}meier}, J.-M. and {Men}, Y.~P. and {Balakrishnan}, V. and {Bezuidenhout}, M.~C.},
        title = "{MeerKAT discovery of 13 new pulsars in Omega Centauri}",
      journal = {\mnras},
         year = 2023,
        month = apr,
       volume = {520},
       number = {3},
        pages = {3847-3856},
          doi = {10.1093/mnras/stad029},
archivePrefix = {arXiv},
       eprint = {2301.03864},
 primaryClass = {astro-ph.HE},
       adsurl = {https://ui.adsabs.harvard.edu/abs/2023MNRAS.520.3847C}
}

@ARTICLE{2025OJAp....854653L,
       author = {{Levin}, Lina and {Bagchi}, Manjari and {Burgay}, Marta and {Deller}, Adam. T. and {Graber}, Vanessa and {Igoshev}, Andrei and {Kramer}, Michael and {Lorimer}, Duncan and {Posselt}, Bettina and {Prabu}, Thiagaraj and {Rajwade}, Kaustubh and {Rea}, Nanda and {Stappers}, Benjamin and {Tauris}, Thomas. M. and {Weltevrede}, Patrick},
        title = "{Understanding the Neutron Star Population with the SKAO telescopes}",
      journal = {The Open Journal of Astrophysics},
         year = 2025,
        month = dec,
       volume = {8},
        pages = {54653},
          doi = {10.33232/001c.154653},
archivePrefix = {arXiv},
       eprint = {2512.16156},
 primaryClass = {astro-ph.HE},
       adsurl = {https://ui.adsabs.harvard.edu/abs/2025OJAp....854653L}
}

@ARTICLE{2025SciBu..70.1568Z,
       author = {{Zhang}, Lei and {Abbate}, Federico and {Li}, Di and {Possenti}, Andrea and {Bailes}, Matthew and {Ridolfi}, Alessandro and {Freire}, Paulo C.~C. and {Ransom}, Scott M. and {Zhang}, Yong-Kun and {Guo}, Meng and {Ni}, Meng-Meng and {Hu}, Jia-Le and {Feng}, Yi and {Wang}, Pei and {Zhang}, Jie and {Zhi}, Qi-Jun},
        title = "{Probing globular cluster with MeerKAT and FAST: a pulsar polarization census}",
      journal = {Science Bulletin},
         year = 2025,
        month = may,
       volume = {70},
       number = {10},
        pages = {1568-1571},
          doi = {10.1016/j.scib.2025.03.022},
archivePrefix = {arXiv},
       eprint = {2503.08291},
 primaryClass = {astro-ph.HE},
       adsurl = {https://ui.adsabs.harvard.edu/abs/2025SciBu..70.1568Z}
}

@ARTICLE{2025OJAp....854251B,
       author = {{Bagchi}, Manjari and {Abbate}, Federico and {Balakrishnan}, Vishnu and {Bernadich}, Miquel Colom i. and {Bhattacharyya}, Bhaswati and {Dutta}, Arunima and {Freire}, Paulo C.~C. and {Halley}, Kriisa and {Hessels}, Jason W.~T. and {Kumari}, Sangeeta and {Lorimer}, Duncan R. and {Possenti}, Andrea and {Nag}, Rouhin and {Ransom}, Scott M. and {Ridolfi}, Alesandro and {Venkatraman Krishnan}, Vivek and {Zhu}, Weiwei},
        title = "{Pulsars in Globular Clusters With the SKAO}",
      journal = {The Open Journal of Astrophysics},
         year = 2025,
        month = dec,
       volume = {8},
        pages = {54251},
          doi = {10.33232/001c.154251},
archivePrefix = {arXiv},
       eprint = {2512.16154},
 primaryClass = {astro-ph.HE},
       adsurl = {https://ui.adsabs.harvard.edu/abs/2025OJAp....854251B}
}

@ARTICLE{2024MNRAS.532.3558K,
       author = {{Karastergiou}, A. and {Johnston}, S. and {Posselt}, B. and {Oswald}, L.~S. and {Kramer}, M. and {Weltevrede}, P.},
        title = "{The Thousand-Pulsar-Array programme on MeerKAT -- XV. A comparison of the radio emission properties of slow and millisecond pulsars}",
      journal = {\mnras},
         year = 2024,
        month = aug,
       volume = {532},
       number = {3},
        pages = {3558-3566},
          doi = {10.1093/mnras/stae1694},
archivePrefix = {arXiv},
       eprint = {2407.06836},
 primaryClass = {astro-ph.HE},
       adsurl = {https://ui.adsabs.harvard.edu/abs/2024MNRAS.532.3558K}
}

@ARTICLE{1997ApJ...482..971C,
       author = {{Cordes}, J.~M. and {Chernoff}, David F.},
        title = "{Neutron Star Population Dynamics. I. Millisecond Pulsars}",
      journal = {\apj},
         year = 1997,
        month = jun,
       volume = {482},
       number = {2},
        pages = {971-992},
          doi = {10.1086/304179},
archivePrefix = {arXiv},
       eprint = {astro-ph/9706162},
 primaryClass = {astro-ph},
       adsurl = {https://ui.adsabs.harvard.edu/abs/1997ApJ...482..971C}
}

@ARTICLE{2023ApJ...944..225L,
       author = {{Lee}, Jongsu and {Hui}, C.~Y. and {Takata}, J. and {Kong}, A.~K.~H. and {Tam}, Pak-Hin Thomas and {Li}, Kwan-Lok and {Cheng}, K.~S.},
        title = "{A Comparison of Millisecond Pulsar Populations between Globular Clusters and the Galactic Field}",
      journal = {\apj},
         year = 2023,
        month = feb,
       volume = {944},
       number = {2},
          eid = {225},
        pages = {225},
          doi = {10.3847/1538-4357/acb5a3},
archivePrefix = {arXiv},
       eprint = {2302.08776},
 primaryClass = {astro-ph.HE},
       adsurl = {https://ui.adsabs.harvard.edu/abs/2023ApJ...944..225L}
}

@ARTICLE{1992ApJ...391..659B,
       author = {{Bailes}, M. and {Kniffen}, D.~A.},
        title = "{Galactic Gamma-Ray Emission from Radio Pulsars}",
      journal = {\apj},
         year = 1992,
        month = jun,
       volume = {391},
        pages = {659},
          doi = {10.1086/171379},
       adsurl = {https://ui.adsabs.harvard.edu/abs/1992ApJ...391..659B}
}

@INPROCEEDINGS{2013IAUS..291..127R,
       author = {{Roberts}, Mallory S.~E.},
        title = "{Surrounded by spiders! New black widows and redbacks in the Galactic field}",
    booktitle = {Neutron Stars and Pulsars: Challenges and Opportunities after 80 years},
         year = 2013,
       editor = {{van Leeuwen}, Joeri},
       series = {IAU Symposium},
       volume = {291},
        month = mar,
        pages = {127-132},
          doi = {10.1017/S174392131202337X},
archivePrefix = {arXiv},
       eprint = {1210.6903},
 primaryClass = {astro-ph.HE},
       adsurl = {https://ui.adsabs.harvard.edu/abs/2013IAUS..291..127R}
}

@ARTICLE{2011ApJ...742...51B,
       author = {{Boyles}, J. and {Lorimer}, D.~R. and {Turk}, P.~J. and {Mnatsakanov}, R. and {Lynch}, R.~S. and {Ransom}, S.~M. and {Freire}, P.~C. and {Belczynski}, K.},
        title = "{Young Radio Pulsars in Galactic Globular Clusters}",
      journal = {\apj},
         year = 2011,
        month = nov,
       volume = {742},
       number = {1},
          eid = {51},
        pages = {51},
          doi = {10.1088/0004-637X/742/1/51},
archivePrefix = {arXiv},
       eprint = {1108.4402},
 primaryClass = {astro-ph.SR},
       adsurl = {https://ui.adsabs.harvard.edu/abs/2011ApJ...742...51B}
}

@ARTICLE{2026A&A...710A.130C,
       author = {{Chen}, W. and {Risbud}, D. and {Freire}, P.~C.~C. and {Ridolfi}, A. and {Barr}, E. and {Kramer}, M. and {Stappers}, B. and {Camilo}, F. and {Abbate}, F. and {Possenti}, A. and {Men}, Y.~P. and {Padmanabh}, P.~V. and {Ransom}, S.~M. and {Vleeschower}, L. and {Venkatraman Krishnan}, V. and {Champion}, D.~J. and {Breton}, R. and {Balakrishnan}, V. and {Buchner}, S.},
        title = "{Fifteen new millisecond pulsars in 47 Tucanae}",
      journal = {\aap},
         year = 2026,
        month = jun,
       volume = {710},
          eid = {A130},
        pages = {A130},
          doi = {10.1051/0004-6361/202659650},
archivePrefix = {arXiv},
       eprint = {2605.06492},
 primaryClass = {astro-ph.HE},
       adsurl = {https://ui.adsabs.harvard.edu/abs/2026A&A...710A.130C}
}

@ARTICLE{Freire2011,
       author = {{Freire}, P.~C.~C. and {Abdo}, A.~A. and {Ajello}, M. and {Allafort}, A. and {Ballet}, J. and {Barbiellini}, G. and {Bastieri}, D. and {Bechtol}, K. and {Bellazzini}, R. and {Bignami}, G.~F. and et al.},
        title = "{Fermi Detection of a Luminous $\gamma$-Ray Pulsar in a Globular Cluster}",
      journal = {Science},
         year = 2011,
        month = nov,
       volume = {334},
       number = {6059},
        pages = {1107},
          doi = {10.1126/science.1211533}
}

@ARTICLE{Wu2013,
       author = {{Wu}, J.~H.~K. and {Kong}, A.~K.~H. and {Huang}, R.~H.~H. and {Cheng}, K.~S. and {Tam}, P.~H.~T. and {Takata}, J. and {Lin}, L.~C.~C. and {Hui}, C.~Y.},
        title = "{Search for Pulsed $\gamma$-Ray Emission from Globular Cluster M28}",
      journal = {\apjl},
         year = 2013,
        month = mar,
       volume = {765},
       number = {2},
          eid = {L47},
        pages = {L47},
          doi = {10.1088/2041-8205/765/2/L47}
}

@ARTICLE{Gautam2022,
       author = {{Gautam}, T. and {Ridolfi}, A. and {Freire}, P.~C.~C. and {Ransom}, S.~M. and {Possenti}, A. and {Stairs}, I.~H. and {Kramer}, M. and {Cadelano}, M. and {Pallanca}, C.},
        title = "{Gamma-ray pulsations from the energetic millisecond pulsar PSR J1835-3259B in the globular cluster NGC 6652}",
      journal = {\aap},
         year = 2022,
       volume = {664},
          eid = {A48},
        pages = {A48},
          doi = {10.1051/0004-6361/202243734}
}

@ARTICLE{Zhang2022,
       author = {{Zhang}, S. and {Hui}, C.~Y. and {Li}, K.~L. and {Kong}, A.~K.~H. and {Takata}, J. and {Cheng}, K.~S.},
        title = "{Detection of gamma-ray pulsations from PSR J1835-3259B in the globular cluster NGC 6652}",
      journal = {\mnras},
         year = 2022,
       volume = {517},
       number = {1},
        pages = {1138},
          doi = {10.1093/mnras/stac265}
}

@ARTICLE{2024arXiv241211271S,
       author = {{Singleton}, Justine and {DeCesar}, Megan and {Dai}, Shi and {Bhakta}, Deven and {Ransom}, Scott and {Strader}, Jay and {Chomiuk}, Laura and {Miller-Jones}, James},
        title = "{Timing of Seven Isolated Pulsars in the Globular Cluster Terzan 1}",
      journal = {arXiv e-prints},
         year = 2024,
        month = dec,
          eid = {arXiv:2412.11271},
        pages = {arXiv:2412.11271},
          doi = {10.48550/arXiv.2412.11271},
archivePrefix = {arXiv},
       eprint = {2412.11271},
 primaryClass = {astro-ph.HE},
       adsurl = {https://ui.adsabs.harvard.edu/abs/2024arXiv241211271S}
}

@ARTICLE{Abdo2010,
       author = {{Abdo}, A.~A. and {Ackermann}, M. and {Ajello}, M. and {Atwood}, W.~B. and {Axelsson}, M. and {Baldini}, L. and {Ballet}, J. and {Barbiellini}, G. and {Bastieri}, D. and {Baughman}, B.~M. and et al.},
        title = "{Fermi Large Area Telescope observations of gamma-ray emission from Galactic globular clusters}",
      journal = {\aap},
         year = 2010,
        month = dec,
       volume = {524},
          eid = {A75},
        pages = {A75},
          doi = {10.1051/0004-6361/201014458},
archivePrefix = {arXiv},
       eprint = {1003.3588},
 primaryClass = {astro-ph.HE},
       adsurl = {https://ui.adsabs.harvard.edu/abs/2010A&A...524A..75A}
}

@ARTICLE{2006MNRAS.369..407C,
       author = {{Cackett}, E.~M. and {Wijnands}, R. and {Heinke}, C.~O. and {Pooley}, D. and {Lewin}, W.~H.~G. and {Grindlay}, J.~E. and {Edmonds}, P.~D. and {Jonker}, P.~G. and {Miller}, J.~M.},
        title = "{A Chandra X-ray observation of the globular cluster Terzan 1}",
      journal = {\mnras},
         year = 2006,
        month = jun,
       volume = {369},
       number = {1},
        pages = {407-415},
          doi = {10.1111/j.1365-2966.2006.10315.x},
archivePrefix = {arXiv},
       eprint = {astro-ph/0512168},
 primaryClass = {astro-ph},
       adsurl = {https://ui.adsabs.harvard.edu/abs/2006MNRAS.369..407C}
}

@ARTICLE{2023arXiv230712546B,
       author = {{Ballet}, J. and {Bruel}, P. and {Burnett}, T.~H. and {Lott}, B. and {The Fermi-LAT collaboration}},
        title = "{Fermi Large Area Telescope Fourth Source Catalog Data Release 4 (4FGL-DR4)}",
      journal = {arXiv e-prints},
         year = 2023,
        month = jul,
          eid = {arXiv:2307.12546},
        pages = {arXiv:2307.12546},
          doi = {10.48550/arXiv.2307.12546},
archivePrefix = {arXiv},
       eprint = {2307.12546},
 primaryClass = {astro-ph.HE},
       adsurl = {https://ui.adsabs.harvard.edu/abs/2023arXiv230712546B}
}

@ARTICLE{2014ApJ...784...59S,
       author = {{Szary}, Andrzej and {Zhang}, Bing and {Melikidze}, George I. and {Gil}, Janusz and {Xu}, Ren-Xin},
        title = "{Radio Efficiency of Pulsars}",
      journal = {\apj},
         year = 2014,
        month = mar,
       volume = {784},
       number = {1},
          eid = {59},
        pages = {59},
          doi = {10.1088/0004-637X/784/1/59},
archivePrefix = {arXiv},
       eprint = {1402.0228},
 primaryClass = {astro-ph.HE},
       adsurl = {https://ui.adsabs.harvard.edu/abs/2014ApJ...784...59S}
}

@ARTICLE{2023ApJ...958..191S,
       author = {{Smith}, D.~A. and {Abdollahi}, S. and {Ajello}, M. and {Bailes}, M. and {Baldini}, L. and {Ballet}, J. and {Baring}, M.~G. and {Bassa}, C. and {Gonzalez}, J. Becerra and {Bellazzini}, R. and {Berretta}, A. and {Bhattacharyya}, B. and {Bissaldi}, E. and {Bonino}, R. and {Bottacini}, E. and {Bregeon}, J. and {Bruel}, P. and {Burgay}, M. and {Burnett}, T.~H. and {Cameron}, R.~A. and {Camilo}, F. and {Caputo}, R. and {Caraveo}, P.~A. and {Cavazzuti}, E. and {Chiaro}, G. and {Ciprini}, S. and {Clark}, C.~J. and {Cognard}, I. and {Corongiu}, A. and {Orestano}, P. Cristarella and {Crnogorcevic}, M. and {Cuoco}, A. and {Cutini}, S. and {D'Ammando}, F. and {de Angelis}, A. and {DeCesar}, M.~E. and {De Gaetano}, S. and {de Menezes}, R. and {Deneva}, J. and {de Palma}, F. and {Di Lalla}, N. and {Dirirsa}, F. and {Di Venere}, L. and {Dom{\'\i}nguez}, A. and {Dumora}, D. and {Fegan}, S.~J. and {Ferrara}, E.~C. and {Fiori}, A. and {Fleischhack}, H. and {Flynn}, C. and {Franckowiak}, A. and {Freire}, P.~C.~C. and {Fukazawa}, Y. and {Fusco}, P. and {Galanti}, G. and {Gammaldi}, V. and {Gargano}, F. and {Gasparrini}, D. and {Giacchino}, F. and {Giglietto}, N. and {Giordano}, F. and {Giroletti}, M. and {Green}, D. and {Grenier}, I.~A. and {Guillemot}, L. and {Guiriec}, S. and {Gustafsson}, M. and {Harding}, A.~K. and {Hays}, E. and {Hewitt}, J.~W. and {Horan}, D. and {Hou}, X. and {Jankowski}, F. and {Johnson}, R.~P. and {Johnson}, T.~J. and {Johnston}, S. and {Kataoka}, J. and {Keith}, M.~J. and {Kerr}, M. and {Kramer}, M. and {Kuss}, M. and {Latronico}, L. and {Lee}, S.-H. and {Li}, D. and {Li}, J. and {Limyansky}, B. and {Longo}, F. and {Loparco}, F. and {Lorusso}, L. and {Lovellette}, M.~N. and {Lower}, M. and {Lubrano}, P. and {Lyne}, A.~G. and {Maan}, Y. and {Maldera}, S. and {Manchester}, R.~N. and {Manfreda}, A. and {Marelli}, M. and {Mart{\'\i}-Devesa}, G. and {Mazziotta}, M.~N. and {McEnery}, J.~E. and {Mereu}, I. and {Michelson}, P.~F. and {Mickaliger}, M. and {Mitthumsiri}, W. and {Mizuno}, T. and {Moiseev}, A.~A. and {Monzani}, M.~E. and {Morselli}, A. and {Negro}, M. and {Nemmen}, R. and {Nieder}, L. and {Nuss}, E. and {Omodei}, N. and {Orienti}, M. and {Orlando}, E. and {Ormes}, J.~F. and {Palatiello}, M. and {Paneque}, D. and {Panzarini}, G. and {Parthasarathy}, A. and {Persic}, M. and {Pesce-Rollins}, M. and {Pillera}, R. and {Poon}, H. and {Porter}, T.~A. and {Possenti}, A. and {Principe}, G. and {Rain{\`o}}, S. and {Rando}, R. and {Ransom}, S.~M. and {Ray}, P.~S. and {Razzano}, M. and {Razzaque}, S. and {Reimer}, A. and {Reimer}, O. and {Renault-Tinacci}, N. and {Romani}, R.~W. and {S{\'a}nchez-Conde}, M. and {Parkinson}, P.~M. Saz and {Scotton}, L. and {Serini}, D. and {Sgr{\`o}}, C. and {Shannon}, R. and {Sharma}, V. and {Shen}, Z. and {Siskind}, E.~J. and {Spandre}, G. and {Spinelli}, P. and {Stappers}, B.~W. and {Stephens}, T.~E. and {Suson}, D.~J. and {Tabassum}, S. and {Tajima}, H. and {Tak}, D. and {Theureau}, G. and {Thompson}, D.~J. and {Tibolla}, O. and {Torres}, D.~F. and {Valverde}, J. and {Venter}, C. and {Wadiasingh}, Z. and {Wang}, N. and {Wang}, N. and {Wang}, P. and {Weltevrede}, P. and {Wood}, K. and {Yan}, J. and {Zaharijas}, G. and {Zhang}, C. and {Zhu}, W.},
        title = "{The Third Fermi Large Area Telescope Catalog of Gamma-Ray Pulsars}",
      journal = {\apj},
         year = 2023,
        month = dec,
       volume = {958},
       number = {2},
          eid = {191},
        pages = {191},
          doi = {10.3847/1538-4357/acee67},
archivePrefix = {arXiv},
       eprint = {2307.11132},
 primaryClass = {astro-ph.HE},
       adsurl = {https://ui.adsabs.harvard.edu/abs/2023ApJ...958..191S}
}

@ARTICLE{2015MNRAS.450.2185L,
       author = {{Lorimer}, D.~R. and {Esposito}, P. and {Manchester}, R.~N. and {Possenti}, A. and {Lyne}, A.~G. and {McLaughlin}, M.~A. and {Kramer}, M. and {Hobbs}, G. and {Stairs}, I.~H. and {Burgay}, M. and {Eatough}, R.~P. and {Keith}, M.~J. and {Faulkner}, A.~J. and {D'Amico}, N. and {Camilo}, F. and {Corongiu}, A. and {Crawford}, F.},
        title = "{The Parkes multibeam pulsar survey - VII. Timing of four millisecond pulsars and the underlying spin-period distribution of the Galactic millisecond pulsar population}",
      journal = {\mnras},
         year = 2015,
        month = jun,
       volume = {450},
       number = {2},
        pages = {2185-2194},
          doi = {10.1093/mnras/stv804},
archivePrefix = {arXiv},
       eprint = {1501.05516},
 primaryClass = {astro-ph.IM},
       adsurl = {https://ui.adsabs.harvard.edu/abs/2015MNRAS.450.2185L}
}

@ARTICLE{2013MNRAS.433..259B,
       author = {{Burgay}, M. and {Bailes}, M. and {Bates}, S.~D. and {Bhat}, N.~D.~R. and {Burke-Spolaor}, S. and {Champion}, D.~J. and {Coster}, P. and {D'Amico}, N. and {Johnston}, S. and {Keith}, M.~J. and {Kramer}, M. and {Levin}, L. and {Lyne}, A.~G. and {Milia}, S. and {Ng}, C. and {Possenti}, A. and {Stappers}, B.~W. and {Thornton}, D. and {Tiburzi}, C. and {van Straten}, W. and {Bassa}, C.~G.},
        title = "{The High Time Resolution Universe Pulsar Survey - VII. Discovery of five millisecond pulsars and the different luminosity properties of binary and isolated recycled pulsars}",
      journal = {\mnras},
         year = 2013,
        month = jul,
       volume = {433},
       number = {1},
        pages = {259-269},
          doi = {10.1093/mnras/stt721},
archivePrefix = {arXiv},
       eprint = {1307.7629},
 primaryClass = {astro-ph.HE},
       adsurl = {https://ui.adsabs.harvard.edu/abs/2013MNRAS.433..259B}
}

@ARTICLE{2024ApJ...974..144B,
       author = {{Berteaud}, Joanna and {Eckner}, Christopher and {Calore}, Francesca and {Clavel}, Ma{\"\i}ca and {Haggard}, Daryl},
        title = "{Simulation-based Inference of Radio Millisecond Pulsars in Globular Clusters}",
      journal = {\apj},
         year = 2024,
        month = oct,
       volume = {974},
       number = {1},
          eid = {144},
        pages = {144},
          doi = {10.3847/1538-4357/ad6b1e},
archivePrefix = {arXiv},
       eprint = {2405.15691},
 primaryClass = {astro-ph.HE},
       adsurl = {https://ui.adsabs.harvard.edu/abs/2024ApJ...974..144B}
}

@ARTICLE{2006ApJ...643..332F,
       author = {{Faucher-Gigu{\`e}re}, Claude-Andr{\'e} and {Kaspi}, Victoria M.},
        title = "{Birth and Evolution of Isolated Radio Pulsars}",
      journal = {\apj},
         year = 2006,
        month = may,
       volume = {643},
       number = {1},
        pages = {332-355},
          doi = {10.1086/501516},
archivePrefix = {arXiv},
       eprint = {astro-ph/0512585},
 primaryClass = {astro-ph},
       adsurl = {https://ui.adsabs.harvard.edu/abs/2006ApJ...643..332F}
}

@ARTICLE{2011MNRAS.418..477B,
       author = {{Bagchi}, Manjari and {Lorimer}, D.~R. and {Chennamangalam}, Jayanth},
        title = "{Luminosities of recycled radio pulsars in globular clusters}",
      journal = {\mnras},
         year = 2011,
        month = nov,
       volume = {418},
       number = {1},
        pages = {477-489},
          doi = {10.1111/j.1365-2966.2011.19498.x},
archivePrefix = {arXiv},
       eprint = {1107.4521},
 primaryClass = {astro-ph.SR},
       adsurl = {https://ui.adsabs.harvard.edu/abs/2011MNRAS.418..477B}
}

@ARTICLE{2013MNRAS.431..874C,
       author = {{Chennamangalam}, Jayanth and {Lorimer}, D.~R. and {Mandel}, Ilya and {Bagchi}, Manjari},
        title = "{Constraining the luminosity function parameters and population size of radio pulsars in globular clusters}",
      journal = {\mnras},
         year = 2013,
        month = may,
       volume = {431},
       number = {1},
        pages = {874-881},
          doi = {10.1093/mnras/stt205},
archivePrefix = {arXiv},
       eprint = {1207.5732},
 primaryClass = {astro-ph.SR},
       adsurl = {https://ui.adsabs.harvard.edu/abs/2013MNRAS.431..874C}
}

@ARTICLE{1987Natur.328..399L,
       author = {{Lyne}, A.~G. and {Brinklow}, A. and {Middleditch}, J. and {Kulkarni}, S.~R. and {Backer}, D.~C.},
        title = "{The discovery of a millisecond pulsar in the globular cluster M28}",
      journal = {\nat},
         year = 1987,
        month = jul,
       volume = {328},
       number = {6129},
        pages = {399-401},
          doi = {10.1038/328399a0},
       adsurl = {https://ui.adsabs.harvard.edu/abs/1987Natur.328..399L}
}

@ARTICLE{1975ApJ...199L.143C,
       author = {{Clark}, G.~W.},
        title = "{X-ray binaries in globular clusters.}",
      journal = {\apjl},
         year = 1975,
        month = aug,
       volume = {199},
        pages = {L143-L145},
          doi = {10.1086/181869},
       adsurl = {https://ui.adsabs.harvard.edu/abs/1975ApJ...199L.143C}
}

@BOOK{2023pbse.book.....T,
       author = {{Tauris}, Thomas M. and {van den Heuvel}, Edward P.~J.},
        title = "{Physics of Binary Star Evolution. From Stars to X-ray Binaries and Gravitational Wave Sources}",
         year = 2023,
          doi = {10.48550/arXiv.2305.09388},
       adsurl = {https://ui.adsabs.harvard.edu/abs/2023pbse.book.....T}
}

@ARTICLE{1982Natur.300..615B,
       author = {{Backer}, D.~C. and {Kulkarni}, S.~R. and {Heiles}, C. and {Davis}, M.~M. and {Goss}, W.~M.},
        title = "{A millisecond pulsar}",
      journal = {\nat},
         year = 1982,
        month = dec,
       volume = {300},
       number = {5893},
        pages = {615-618},
          doi = {10.1038/300615a0},
       adsurl = {https://ui.adsabs.harvard.edu/abs/1982Natur.300..615B}
}

@ARTICLE{2014ApJ...793...97K,
       author = {{Kalapotharakos}, Constantinos and {Harding}, Alice K. and {Kazanas}, Demosthenes},
        title = "{Gamma-Ray Emission in Dissipative Pulsar Magnetospheres: From Theory to Fermi Observations}",
      journal = {\apj},
         year = 2014,
        month = oct,
       volume = {793},
       number = {2},
          eid = {97},
        pages = {97},
          doi = {10.1088/0004-637X/793/2/97},
archivePrefix = {arXiv},
       eprint = {1310.3545},
 primaryClass = {astro-ph.HE},
       adsurl = {https://ui.adsabs.harvard.edu/abs/2014ApJ...793...97K}
}

@ARTICLE{baumgardt2021,
       author = {{Baumgardt}, H. and {Vasiliev}, E.},
        title = "{Accurate distances to Galactic globular clusters through a combination of Gaia EDR3, HST, and literature data}",
      journal = {\mnras},
         year = 2021,
        month = aug,
       volume = {505},
       number = {4},
        pages = {5957-5977},
          doi = {10.1093/mnras/stab1474},
archivePrefix = {arXiv},
       eprint = {2105.09526},
 primaryClass = {astro-ph.GA},
       adsurl = {https://ui.adsabs.harvard.edu/abs/2021MNRAS.505.5957B}
}

@ARTICLE{shawaiz2025,
       author = {{Tabassum}, Shawaiz and {Lorimer}, Duncan R.},
        title = "{Monte Carlo Evaluations of Gamma-Ray and Radio Pulsar Populations}",
      journal = {\apj},
         year = 2025,
        month = jul,
       volume = {988},
       number = {1},
          eid = {78},
        pages = {78},
          doi = {10.3847/1538-4357/ade13f},
archivePrefix = {arXiv},
       eprint = {2504.02677},
 primaryClass = {astro-ph.HE},
       adsurl = {https://ui.adsabs.harvard.edu/abs/2025ApJ...988...78T}
}

@ARTICLE{holst2025,
       author = {{Holst}, Ian and {Hooper}, Dan},
        title = "{New determination of the millisecond pulsar gamma-ray luminosity function and implications for the Galactic Center gamma-ray excess}",
      journal = {\prd},
         year = 2025,
        month = jan,
       volume = {111},
       number = {2},
          eid = {023048},
        pages = {023048},
          doi = {10.1103/PhysRevD.111.023048},
archivePrefix = {arXiv},
       eprint = {2403.00978},
 primaryClass = {astro-ph.HE},
       adsurl = {https://ui.adsabs.harvard.edu/abs/2025PhRvD.111b3048H}
}

@ARTICLE{gonthier2018,
       author = {{Gonthier}, Peter L. and {Harding}, Alice K. and {Ferrara}, Elizabeth C. and {Frederick}, Sara E. and {Mohr}, Victoria E. and {Koh}, Yew-Meng},
        title = "{Population Syntheses of Millisecond Pulsars from the Galactic Disk and Bulge}",
      journal = {\apj},
         year = 2018,
        month = aug,
       volume = {863},
       number = {2},
          eid = {199},
        pages = {199},
          doi = {10.3847/1538-4357/aad08d},
archivePrefix = {arXiv},
       eprint = {1806.11215},
 primaryClass = {astro-ph.HE},
       adsurl = {https://ui.adsabs.harvard.edu/abs/2018ApJ...863..199G}
}

@ARTICLE{kalap2019,
       author = {{Kalapotharakos}, Constantinos and {Harding}, Alice K. and {Kazanas}, Demosthenes and {Wadiasingh}, Zorawar},
        title = "{A Fundamental Plane for Gamma-Ray Pulsars}",
      journal = {\apjl},
         year = 2019,
        month = sep,
       volume = {883},
       number = {1},
          eid = {L4},
        pages = {L4},
          doi = {10.3847/2041-8213/ab3e0a},
archivePrefix = {arXiv},
       eprint = {1904.01765},
 primaryClass = {astro-ph.HE},
       adsurl = {https://ui.adsabs.harvard.edu/abs/2019ApJ...883L...4K}
}

@ARTICLE{kalap2022,
       author = {{Kalapotharakos}, Constantinos and {Wadiasingh}, Zorawar and {Harding}, Alice K. and {Kazanas}, Demosthenes},
        title = "{The Fundamental Plane Relation for Gamma-Ray Pulsars Implied by 4FGL}",
      journal = {\apj},
         year = 2022,
        month = jul,
       volume = {934},
       number = {1},
          eid = {65},
        pages = {65},
          doi = {10.3847/1538-4357/ac78e3},
archivePrefix = {arXiv},
       eprint = {2203.13276},
 primaryClass = {astro-ph.HE},
       adsurl = {https://ui.adsabs.harvard.edu/abs/2022ApJ...934...65K}
}

@BOOK{lorimer2004,
       author = {{Lorimer}, D.~R. and {Kramer}, M.},
        title = "{Handbook of Pulsar Astronomy}",
         year = 2004,
       volume = {4},
       adsurl = {https://ui.adsabs.harvard.edu/abs/2004hpa..book.....L}
}

@ARTICLE{kramer1998,
       author = {{Kramer}, Michael and {Xilouris}, Kiriaki M. and {Lorimer}, Duncan R. and {Doroshenko}, Oleg and {Jessner}, Axel and {Wielebinski}, Richard and {Wolszczan}, Alexander and {Camilo}, Fernando},
        title = "{The Characteristics of Millisecond Pulsar Emission. I. Spectra, Pulse Shapes, and the Beaming Fraction}",
      journal = {\apj},
         year = 1998,
        month = jul,
       volume = {501},
       number = {1},
        pages = {270-285},
          doi = {10.1086/305790},
archivePrefix = {arXiv},
       eprint = {astro-ph/9801177},
 primaryClass = {astro-ph},
       adsurl = {https://ui.adsabs.harvard.edu/abs/1998ApJ...501..270K}
}

@ARTICLE{bahramian2013,
       author = {{Bahramian}, Arash and {Heinke}, Craig O. and {Sivakoff}, Gregory R. and {Gladstone}, Jeanette C.},
        title = "{Stellar Encounter Rate in Galactic Globular Clusters}",
      journal = {\apj},
         year = 2013,
        month = apr,
       volume = {766},
       number = {2},
          eid = {136},
        pages = {136},
          doi = {10.1088/0004-637X/766/2/136},
archivePrefix = {arXiv},
       eprint = {1302.2549},
 primaryClass = {astro-ph.HE},
       adsurl = {https://ui.adsabs.harvard.edu/abs/2013ApJ...766..136B}
}

@ARTICLE{turk2013,
       author = {{Turk}, P.~J. and {Lorimer}, D.~R.},
        title = "{An empirical Bayesian analysis applied to the globular cluster pulsar population}",
      journal = {\mnras},
         year = 2013,
        month = dec,
       volume = {436},
       number = {4},
        pages = {3720-3726},
          doi = {10.1093/mnras/stt1850},
archivePrefix = {arXiv},
       eprint = {1309.7317},
 primaryClass = {astro-ph.GA},
       adsurl = {https://ui.adsabs.harvard.edu/abs/2013MNRAS.436.3720T}
}

@ARTICLE{hui2010,
       author = {{Hui}, C.~Y. and {Cheng}, K.~S. and {Taam}, Ronald E.},
        title = "{Dynamical Formation of Millisecond Pulsars in Globular Clusters}",
      journal = {\apj},
         year = 2010,
        month = may,
       volume = {714},
       number = {2},
        pages = {1149-1154},
          doi = {10.1088/0004-637X/714/2/1149},
archivePrefix = {arXiv},
       eprint = {1003.4332},
 primaryClass = {astro-ph.HE},
       adsurl = {https://ui.adsabs.harvard.edu/abs/2010ApJ...714.1149H}
}

\begin{table*}
	\centering
	\caption{The sample of 118 GCs modeled in this work. For each GC, we list the Galactic latitude $l$ and longitude $b$, distance $D$ \citep[from][]{baumgardt2021}, number of observed radio pulsars $N_{\rm PSR, R}$ if any (from Freire's catalogue), measured or upper-limit gamma-ray fluxes $F_{\gamma}$ (37 fluxes from {\it Fermi} LAT detections, 63 upper limits from {\it Fermi} LAT sensitivities, 15 upper limits from 4FGL Objects, and 3 from bright MSP detections), closest 4FGL object for GCs that use their fluxes as upper limits, lowest known flux density $S_{\rm min}$ of a detected pulsar scaled to the 1400~MHz band, and stellar encounter rate $\Gamma$ \citep[derived by][]{bahramian2013}. All upper limit fluxes are denoted by $<$, and 4FGL fluxes are denoted by *. For GCs without any detectable pulsars, we adopt a fiducial $S_{\rm min}$ of 0.1 mJy, denoted by $\dagger$. 
    }
	\label{tab:Table_1}
    \begin{tabular}{lrrllllllll}
\toprule
GC & \multicolumn{1}{c}{$l$} & \multicolumn{1}{c}{$b$} & \multicolumn{1}{c}{$D$} & $N_{\rm PSR, R}$ & $F_{\gamma}$ & 4FGL Object & $S_{\rm min}$ & $\Gamma$ & $\Gamma_{\rm lower}$  & $\Gamma_{\rm upper}$ \\
 & \multicolumn{1}{c}{(deg)} & \multicolumn{1}{c}{(deg)} & \multicolumn{1}{c}{(kpc)} &  & ($10^{-15} \text{ W m}^{-2}$) &  & (mJy) & \\\midrule
Ter 5 & 3.84 & 1.69 & 6.62 & 49 & 80 $\pm{}$ 3 &  & 0.007 & 6800 & 3780 & 7840\\
NGC 7078 & 65.01 & $-$27.31 & 10.7 & 15 & 3.6 $\pm{}$ 0.6 &  & 0.018 & 4510 & 3520 & 5870\\
NGC 6715 & 5.61 & $-$14.09 & 26.3 &  & 2.7 $\pm{}$ 0.6 &  &  & 2520 & 2250 & 2750\\
Ter 6 & 358.57 & $-$2.16 & 7.27 & 1 & 6 $\pm{}$ 1 &  & 0.030 & 2470 & 753 & 7540\\
NGC 6441 & 353.53 & $-$5.01 & 12.7 & 9 & 16 $\pm{}$ 1 &  & 0.018 & 2300 & 1660 & 3270\\
NGC 6266 & 353.57 & 7.32 & 6.41 & 10 & 20 $\pm{}$ 1 &  & 0.009 & 1670 & 1100 & 2380\\
NGC 1851 & 244.51 & $-$35.04 & 12.0 & 15 & 1.9 $\pm{}$ 0.3 &  & 0.011 & 1530 & 1340 & 1730\\
NGC 6440 & 7.73 & 3.80 & 8.25 & 8 & 15 $\pm{}$ 2 &  & 0.03 & 1400 & 923 & 2030\\
NGC 6624 & 2.79 & $-$7.91 & 8.02 & 12 & $13.9\pm1.8$ & & 0.031 & 1150 & 972 & 1260\\
NGC 6681 & 2.85 & $-$12.51 & 9.36 & 3 & 2.0 $\pm{}$ 0.6 &  & 0.1$^\dagger$ & 1040 & 848 & 1310\\
NGC 104 & 305.90 & $-$44.89 & 4.52 & 42 & 27.4 $\pm{}$ 0.6 &  & 0.013 & 1000 & 866 & 1150\\
NGC 5824 & 332.55 & 22.07 & 31.7 &  & <0.7 &  &  & 984 & 829 & 1160\\
Pal 2 & 170.53 & $-$9.07 & 26.2 &  & <0.8 &  &  & 929 & 374 & 1770\\
NGC 2808 & 282.19 & $-$11.25 & 10.1 & 4 & 3.9 $\pm{}$ 0.5 &  &  & 923 & 840 & 990\\
NGC 6388 & 345.56 & $-$6.74 & 11.2 &  & 19 $\pm{}$ 1 &  &  & 899 & 686 & 1140\\
NGC 6293 & 357.62 & 7.83 & 9.19 &  & <2 &  & 0.1$^\dagger$ & 847 & 608 & 1220\\
NGC 362 & 301.53 & $-$46.25 & 8.83 & 12 & 1.5 $\pm{}$ 0.3 &  &  & 735 & 618 & 872\\
NGC 6652 & 1.53 & $-$11.38 & 9.46 & 2 & $7.2\pm0.7$ & &  & 700 & 511 & 992\\
NGC 6284 & 358.35 & 9.94 & 14.2 &  & <1 &  &  & 666 & 561 & 788\\
NGC 6626 & 7.80 & $-$5.58 & 5.37 & 14 & $31.7\pm2.9$ & & 0.011 & 648 & 557 & 732\\
NGC 6093 & 352.67 & 19.46 & 10.3 & 3 & 3.8 $\pm{}$ 0.7 &  &  & 532 & 463 & 591\\
NGC 7089 & 53.37 & $-$35.77 & 11.7 & 10 & <0.6 &  &  & 518 & 447 & 596\\
NGC 5286 & 311.61 & 10.57 & 11.1 &  & 2.6 $\pm{}$ 0.6 &  &  & 458 & 397 & 516\\
NGC 6752 & 336.49 & $-$25.63 & 4.12 & 9 & 3.6 $\pm{}$ 0.4 &  & 0.05 & 401 & 275 & 583\\
NGC 6541 & 349.29 & $-$11.19 & 7.61 &  & 4.0 $\pm{}$ 0.7 &  &  & 386 & 323 & 481\\
NGC 6453 & 355.72 & $-$3.87 & 10.1 &  & <7* & J1751.1--3455 &  & 371 & 282 & 499\\
NGC 6522 & 1.02 & $-$3.93 & 7.29 & 6 & <4* & J1804.9--3001 & 0.004 & 363 & 264 & 476\\
NGC 6517 & 19.22 & 6.76 & 9.23 & 21 & <2 &  & 0.002 & 338 & 240 & 490\\
NGC 7099 & 27.18 & $-$46.84 & 8.46 & 2 & <0.6 &  & 0.08 & 324 & 243 & 448\\
NGC 6139 & 342.36 & 6.94 & 10.0 &  & 5 $\pm{}$ 1 &  &  & 307 & 218 & 401\\
NGC 6864 & 20.30 & $-$25.75 & 20.5 &  & <0.6 &  &  & 307 & 225 & 402\\
NGC 6528 & 1.14 & $-$4.17 & 7.83 &  & 3.5 $\pm{}$ 0.9 &  &  & 278 & 229 & 392\\
NGC 6341 & 68.34 & 34.86 & 8.50 & 2 & 1.1 $\pm{}$ 0.2 &  &  & 270 & 241 & 300\\
NGC 6273 & 356.87 & 9.38 & 8.34 &  & <6* & J1701.0--2617 &  & 200 & 161 & 267\\
NGC 5272 & 42.22 & 78.71 & 10.2 & 5 & <0.5 &  & 0.006 & 194 & 176 & 227\\
NGC 5694 & 331.06 & 30.36 & 34.8 &  & <0.5 &  &  & 191 & 157 & 243\\
NGC 6256 & 347.79 & 3.31 & 7.24 &  & <2 &  &  & 169 & 109 & 288\\
NGC 5904 & 3.86 & 46.80 & 7.48 & 7 & 1.6 $\pm{}$ 0.4 &  & 0.007 & 164 & 134 & 203\\
NGC 6638 & 7.90 & $-$7.15 & 9.78 &  & <4* & J1831.8--2514 &  & 137 & 110 & 176\\
NGC 5946 & 327.58 & 4.19 & 9.64 &  & <1 &  &  & 134 & 89 & 168\\
NGC 6333 & 5.54 & 10.71 & 8.30 &  & <2 &  &  & 131 & 89 & 190\\
NGC 6402 & 21.32 & 14.80 & 9.14 & 5 & 3.5 $\pm{}$ 0.7 &  &  & 124 & 94 & 156\\
NGC 6304 & 355.83 & 5.38 & 6.15 &  & 4 $\pm{}$ 1 &  &  & 123 & 101 & 177\\
NGC 6325 & 0.97 & 8.00 & 7.53 &  & <1 &  &  & 118 & 72 & 163\\
NGC 1904 & 227.23 & $-$29.35 & 13.1 & 1 & 3.0 $\pm{}$ 0.5 &  &  & 116 & 102 & 135\\
NGC 6380 & 350.18 & $-$3.42 & 9.61 &  & <1 &  &  & 116 & 71 & 184\\
NGC 6544 & 5.84 & $-$2.2 & 2.58 & 3 & 8 $\pm{}$ 2 &  & 1.1 & 111 & 74 & 179\\
NGC 6558 & 0.20 & $-$6.02 & 7.47 &  & <1 &  &  & 105 & 86 & 131\\
NGC 6355 & 359.59 & 5.43 & 8.65 &  & <2 &  &  & 99 & 74 & 140\\
NGC 6642 & 9.81 & $-$6.44 & 8.05 &  & <2 &  &  & 98 & 73 & 129\\
\bottomrule
\\
\end{tabular}

\end{table*}

\begin{table*}
	\centering
        \ContinuedFloat
	\caption{Continued.}
	%\label{tab:Table_1}
	\begin{tabular}{lrrllllllll}
\toprule
GC & \multicolumn{1}{c}{$l$} & \multicolumn{1}{c}{$b$} & \multicolumn{1}{c}{$D$} & $N_{\rm PSR, R}$ & $F_{\gamma}$ & 4FGL Object & $S_{\rm min}$ & $\Gamma$ & $\Gamma_{\rm lower}$  & $\Gamma_{\rm upper}$ \\
 & \multicolumn{1}{c}{(deg)} & \multicolumn{1}{c}{(deg)} & \multicolumn{1}{c}{(kpc)} &  & ($10^{-15} \text{ W m}^{-2}$) &  & (mJy) & \\\midrule
NGC 5139 & 309.10 & 14.97 & 5.43 & 19 & 9.1 $\pm{}$ 0.7 &  & 0.019 & 90 & 70 & 117\\
NGC 6637 & 1.72 & $-$10.27 & 8.90 & 2 & <4* & J1830.7--3219 &  & 90 & 72 & 126\\
NGC 6356 & 6.72 & 10.22 & 15.7 &  & <1 &  &  & 88 & 74 & 108\\
NGC 6397 & 338.17 & $-$11.96 & 2.48 & 2 & 3.3 $\pm{}$ 0.8 &  & 0.1$^\dagger$ & 84 & 66 & 102\\
NGC 6656 & 9.89 & $-$7.55 & 3.30 & 4 & 6 $\pm{}$ 1 &  & 0.04 & 78 & 52 & 109\\
NGC 6316 & 357.18 & 5.76 & 11.2 & 1 & 11 $\pm{}$ 1 &  &  & 77 & 62 & 102\\
NGC 6553 & 5.25 & $-$3.02 & 5.33 &  & <2 &  &  & 69 & 50 & 96\\
NGC 6205 & 59.01 & 40.91 & 7.42 & 9 & 0.6 $\pm{}$ 0.1 &  & 0.01 & 69 & 54 & 87\\
NGC 5927 & 326.60 & 4.86 & 8.27 &  & <1 &  &  & 68 & 58 & 81\\
NGC 5986 & 337.02 & 13.27 & 10.5 & 1 & <0.9 &  & 0.029 & 62 & 52 & 78\\
Pal 10 & 52.44 & 2.72 & 8.94 &  & <2 &  &  & 59 & 24 & 102\\
NGC 6760 & 36.11 & $-$3.92 & 8.41 & 2 & <1 &  &  & 57 & 38 & 84\\
NGC 6569 & 0.48 & $-$6.68 & 10.5 &  & <6* & J1812.8--3144 &  & 54 & 33 & 84\\
NGC 6229 & 73.64 & 40.31 & 30.1 &  & <0.5 &  &  & 48 & 38 & 79\\
NGC 6342 & 4.90 & 9.72 & 8.01 & 2 & <6* & J1720.8--1937 & 0.3 & 45 & 32 & 59\\
NGC 6401 & 3.45 & 3.98 & 8.06 &  & 9 $\pm{}$ 2 &  &  & 44 & 33 & 55\\
NGC 6539 & 20.80 & 6.78 & 8.16 & 1 & <10* & J1805.8--0803 & 1.0 & 42 & 27 & 71\\
NGC 6717 & 12.88 & $-$10.9 & 7.52 &  & 3.1 $\pm{}$ 0.5 &  &  & 40 & 26 & 62\\
NGC 6287 & 0.13 & 11.02 & 7.93 &  & <1 &  &  & 36 & 29 & 44\\
NGC 5024 & 332.96 & 79.76 & 18.5 & 5 & <0.4 &  &  & 35 & 26 & 48\\
NGC 6254 & 15.14 & 23.08 & 5.07 & 2 & <3* & J1656.4--0410 &  & 31 & 27 & 36\\
NGC 6712 & 25.35 & $-$4.32 & 7.38 & 1 & 4 $\pm{}$ 1 &  & 0.0161 & 31 & 24 & 36\\
NGC 6934 & 52.10 & $-$18.89 & 15.7 &  & <0.7 &  &  & 30 & 22 & 42\\
NGC 6779 & 62.66 & 8.34 & 10.4 &  & <0.9 &  &  & 28 & 18 & 40\\
NGC 6121 & 350.97 & 15.97 & 1.85 & 1 & <1 &  & 1.6 & 27 & 17 & 38\\
Ter 2 & 356.32 & 2.30 & 7.75 &  & 8 $\pm{}$ 2 &  &  & 22 & 8 & 51\\
Pal 11 & 31.81 & $-$15.58 & 14.0 &  & <1 &  &  & 21 & 14 & 32\\
NGC 5634 & 342.21 & 49.26 & 26.0 &  & <0.5 &  &  & 20 & 13 & 34\\
NGC 4147 & 252.85 & 77.19 & 18.5 &  & <7* & J1209.8+1810 &  & 17 & 10 & 29\\
Pal 6 & 2.09 & 1.78 & 7.05 &  & 4 $\pm{}$ 2 &  &  & 16 & 8 & 29\\
NGC 1261 & 270.54 & $-$52.12 & 16.4 &  & <0.4 &  &  & 15 & 11 & 26\\
NGC 6218 & 15.72 & 26.31 & 5.11 & 2 & 1.4 $\pm{}$ 0.3 &  &  & 13 & 9 & 18\\
NGC 6584 & 342.14 & $-$16.41 & 13.6 &  & <0.5 &  &  & 12 & 8 & 17\\
NGC 6723 & 0.07 & $-$17.3 & 8.27 &  & <0.9 &  &  & 11 & 7 & 19\\
IC 1276 & 21.83 & 5.67 & 4.55 &  & <1 &  &  & 8 & 4 & 16\\
NGC 3201 & 277.23 & 8.64 & 4.74 &  & <0.7 &  &  & 7 & 5 & 11\\
NGC 6171 & 3.37 & 23.01 & 5.63 &  & <1 &  &  & 7 & 5 & 9\\
NGC 6352 & 341.42 & $-$7.17 & 5.54 &  & <0.9 &  &  & 7 & 5 & 8\\
NGC 4590 & 299.63 & 36.05 & 10.4 &  & <2 &  &  & 6 & 4 & 9\\
NGC 6235 & 358.92 & 13.52 & 11.9 &  & <0.9 &  &  & 6 & 4 & 8\\
NGC 6366 & 18.41 & 16.04 & 3.44 &  & <2 &  &  & 5 & 3 & 8\\
NGC 6981 & 35.16 & $-$32.68 & 16.7 &  & <0.7 &  &  & 5 & 3 & 7\\
NGC 6362 & 325.55 & $-$17.57 & 7.65 &  & <0.8 &  &  & 5 & 4 & 6\\
NGC 2298 & 245.63 & $-$16.01 & 9.83 &  & <2* & J0648.6--3623 &  & 4 & 3 & 6\\
Ton 2 & 193.27 & 21.34 & 6.99 &  & <0.7 &  &  & 4 & 3 & 8\\
Pal 8 & 14.10 & $-$6.8 & 11.3 &  & <1 &  &  & 4 & 3 & 7\\
NGC 6809 & 8.79 & $-$23.27 & 5.35 &  & <0.7 &  &  & 3 & 2 & 5\\
NGC 6144 & 351.93 & 15.70 & 8.15 &  & <1 &  &  & 3 & 2 & 4\\
Ter 9 & 3.60 & $-$1.99 & 5.77 &  & <9* & J1802.1--2652 &  & 2 & 1 & 3\\
Ter 7 & 3.39 & $-$20.07 & 24.3 &  & <0.9 &  &  & 2 & 1 & 3\\
\bottomrule
\\
\end{tabular}
	
\end{table*}

\begin{table*}
	\centering
        \ContinuedFloat
	\caption{Continued.}
	%\label{tab:Table_1}
	\begin{tabular}{lrrllllllll}
\toprule
GC & \multicolumn{1}{c}{$l$} & \multicolumn{1}{c}{$b$} & \multicolumn{1}{c}{$D$} & $N_{\rm PSR, R}$ & $F_{\gamma}$ & 4FGL Object & $S_{\rm min}$ & $\Gamma$ & $\Gamma_{\rm lower}$  & $\Gamma_{\rm upper}$ \\
 & \multicolumn{1}{c}{(deg)} & \multicolumn{1}{c}{(deg)} & \multicolumn{1}{c}{(kpc)} &  & ($10^{-15} \text{ W m}^{-2}$) &  & (mJy) & \\\midrule
NGC 6426 & 28.09 & 16.23 & 20.7 &  & <2 &  &  & 2 & 1 & 3\\
NGC 6101 & 317.75 & $-$15.82 & 14.4 &  & <4* & J1622.2--7202 &  & 1.0 & 0.7 & 1.5\\
Pal 1 & 130.06 & 19.03 & 11.2 &  & <0.7 &  &  & 0.9 & 0.7 & 1.5\\
NGC 5897 & 342.95 & 30.29 & 12.5 &  & <0.8 &  &  & 0.9 & 0.7 & 1.2\\
IC 4499 & 307.35 & $-$20.47 & 18.9 &  & <0.7 &  &  & 0.8 & 0.5 & 1.2\\
NGC 288 & 151.29 & $-$89.38 & 8.99 &  & <0.4 &  &  & 0.8 & 0.6 & 1.1\\
HP 1 & 357.43 & 2.12 & 7.00 &  & <20* & J1731.6--3002 &  & 0.7 & 0.4 & 1.1\\
NGC 6496 & 348.03 & $-$10.01 & 9.64 &  & <1 &  &  & 0.7 & 0.4 & 1.3\\
Pal 12 & 30.51 & $-$47.68 & 18.5 &  & <0.8 &  &  & 0.4 & 0.2 & 0.8\\
NGC 6535 & 27.18 & 10.44 & 6.36 &  & <1 &  &  & 0.4 & 0.2 & 0.8\\
Ter 1 & 357.57 & 1.00 & 5.67 & 8 & 7.6 $\pm{}$ 0.8 &  &  & 0.3 & 0.1 & 0.6\\
NGC 5466 & 42.15 & 73.59 & 16.1 &  & <0.4 &  &  & 0.2 & 0.2 & 0.3\\
NGC 4372 & 300.99 & $-$9.88 & 5.71 &  & <1 &  &  & 0.2 & 0.1 & 0.6\\
NGC 7492 & 53.39 & $-$63.48 & 24.4 &  & <0.4 &  &  & 0.2 & 0.1 & 0.4\\
NGC 5053 & 335.70 & 78.95 & 17.5 &  & <0.4 &  &  & 0.1 & 0.1 & 0.2\\
Arp 2 & 8.55 & $-$20.79 & 28.7 &  & <0.7 &  &  & 0.005 & 0.003 & 0.008\\
Pal 5 & 0.85 & 45.86 & 21.9 &  & <6* & J1516.5+0015 &  & 0.002 & 0.001 & 0.003\\
Pal 13 & 87.10 & $-$42.7 & 23.5 &  & <0.8 &  &  & 0.001 & 0.000 & 0.003\\
\bottomrule
\\
\end{tabular}

\end{table*}

\begin{table*} 
	\centering
        \renewcommand{\arraystretch}{1.25} % Default value: 1
	\caption{MSP population estimates for 118 GCs. For each GC, we list the median number of gamma-ray pulsars and 90\% confidence intervals for both models ($N_{\gamma}^A$ and $N_{\gamma}^B$), the mean predicted pulsar abundances ($N_{\Gamma}^A$ and $N_{\Gamma}^B$) and their 1$\sigma$ uncertainties derived from the encounter rate scaling ($N \propto \Gamma^{2/3}$), the corresponding median predicted GC gamma-ray fluxes and 90\% confidence intervals for each model ($F_\gamma^A$ and $F_\gamma^B$), the probability of the GC flux being dominated by a single pulsar ($p_1^A$ and $p_1^B$), the predicted number of currently observable radio pulsars ($N_{\text{Radio}}^A$ and $N_{\text{Radio}}^B$), and the currently known number of radio pulsars ($N_{\rm PSR,R}$).
    For GCs without a {\it Fermi} LAT detection, the quoted $N_\gamma$ values represent upper bounds on the MSP population inferred from flux upper limits, and are denoted by a $<$ symbol.
    }
    \label{tab:Table_2}
	\begin{tabular}{llllllllllll}
\hline
GC & $N^{A}_{\gamma}$ & $N^{B}_{\gamma}$ & $N_{\Gamma}^A$ & $N_{\Gamma}^B$ & $F_{\gamma}^{A}$ & $F_{\gamma}^{B}$ & $p_{b}^{A}$ & $p_{b}^{B}$ & $N^{A}_{\rm Radio}$ & $N^{B}_{\rm Radio}$ & $N_{\rm PSR, R}$\\ &  &  &  &  & \multicolumn{2}{c}{($10^{-12}$ erg s$^{-1}$ cm$^{-2}$)} & & & \multicolumn{2}{c}{($> S_{\rm min}$)} &  \\\hline
Ter 5 & $100_{-28}^{+27}$ & $770_{-390}^{+300}$ & 25 $\pm{}$ 5 & 180 $\pm{}$ 40 & $19_{-9}^{+16}$ & $16_{-9}^{+23}$ & $<0.1\%$ & $0.4\%$ & 18 & 37 & 49\\
NGC 7078 & $13_{-7}^{+10}$ & $110_{-77}^{+81}$ & 20 $\pm{}$ 4 & 140 $\pm{}$ 26 & $6_{-3}^{+5}$ & $5_{-3}^{+9}$ & $7.3\%$ & $12.1\%$ & 7 & 8 & 15\\
NGC 6715 & $55_{-25}^{+29}$ & $420_{-260}^{+280}$ & 15 $\pm{}$ 1 & 110 $\pm{}$ 7 & $0.7_{-0.3}^{+0.7}$ & $0.6_{-0.3}^{+1.1}$ & $0.1\%$ & $0.9\%$ &  &  & \\
Ter 6 & $10_{-7}^{+8}$ & $78_{-61}^{+71}$ & 15 $\pm{}$ 14 & 100 $\pm{}$ 100 & $11_{-10}^{+17}$ & $10_{-9}^{+22}$ & $14.0\%$ & $18.2\%$ & 2 & 3 & 1\\
NGC 6441 & $79_{-26}^{+24}$ & $600_{-300}^{+270}$ & 14 $\pm{}$ 4 & 100 $\pm{}$ 25 & $3_{-2}^{+3}$ & $2_{-1}^{+5}$ & $<0.1\%$ & $0.7\%$ & 4 & 4 & 9\\
NGC 6266 & $26_{-12}^{+12}$ & $210_{-140}^{+110}$ & 12 $\pm{}$ 3 & $85 \pm{}$ 23 & $9_{-6}^{+10}$ & $7_{-5}^{+17}$ & $0.6\%$ & $2.8\%$ & 8 & 16 & 10\\
NGC 1851 & $9_{-6}^{+7}$ & $77_{-60}^{+67}$ & 12 $\pm{}$ 1 & $81 \pm{}$ 7 & $2_{-1}^{+3}$ & $2_{-1}^{+5}$ & $11.0\%$ & $15.0\%$ & 5 & 6 & 15\\
NGC 6440 & $31_{-15}^{+15}$ & $240_{-150}^{+150}$ & 11 $\pm{}$ 3 & $77 \pm{}$ 21 & $5_{-4}^{+7}$ & $4_{-3}^{+10}$ & $0.1\%$ & $2.2\%$ & 4 & 5 & 8\\
NGC 6624 & $28_{-14}^{+15}$ & $210_{-140}^{+140}$ & 10 $\pm{}$ 1 & $70 \pm{}$ 6 & $5_{-3}^{+5}$ & $4_{-2}^{+8}$ & $0.3\%$ & $3.1\%$ & 4 & 4 & 12\\
NGC 6681 & $7_{-5}^{+6}$ & $52_{-40}^{+53}$ & 9 $\pm{}$ 1 & $66 \pm{}$ 10 & $3_{-2}^{+4}$ & $3_{-2}^{+7}$ & $18.3\%$ & $22.0\%$ & 1 & 1 & 3\\
NGC 104 & $18_{-10}^{+9}$ & $140_{-97}^{+91}$ & 9 $\pm{}$ 1 & $65 \pm{}$ 6 & $14_{-8}^{+18}$ & $11_{-6}^{+27}$ & $1.5\%$ & $4.0\%$ & 7 & 15 & 42\\
NGC 5824 & $<34$ & $<290$ & 9 $\pm{}$ 1 & $64 \pm{}$ 8 & $0.3_{-0.2}^{+0.4}$ & $0.2_{-0.1}^{+0.6}$ & $1.1\%$ & $2.5\%$ &  &  & \\
Pal 2 & $<29$ & $<240$ & 9 $\pm{}$ 5 & $63 \pm{}$ 33 & $0.4_{-0.4}^{+0.7}$ & $0.3_{-0.3}^{+0.8}$ & $1.4\%$ & $4.7\%$ &  &  & \\
NGC 2808 & $13_{-8}^{+8}$ & $100_{-79}^{+77}$ & 9 $\pm{}$ 1 & $62 \pm{}$ 4 & $3_{-1}^{+4}$ & $2_{-1}^{+5}$ & $3.4\%$ & $7.5\%$ &  & 1 & 4\\
NGC 6388 & $70_{-24}^{+21}$ & $530_{-280}^{+230}$ & 9 $\pm{}$ 2 & $62 \pm{}$ 11 & $2_{-1}^{+3}$ & $2_{-1}^{+4}$ & $<0.1\%$ & $0.4\%$ &  &  & \\
NGC 6293 & $<9$ & $<79$ & 9 $\pm{}$ 2 & $60 \pm{}$ 15 & $3_{-2}^{+4}$ & $2_{-2}^{+6}$ & $32.2\%$ & $25.9\%$ &  &  & \\
NGC 362 & $5_{-4}^{+4}$ & $36_{-29}^{+38}$ & 8 $\pm{}$ 1 & $55 \pm{}$ 7 & $3_{-2}^{+4}$ & $2_{-1}^{+6}$ & $32.6\%$ & $26.0\%$ & 1 & 1 & 12\\
NGC 6652 & $21_{-12}^{+10}$ & $160_{-110}^{+110}$ & 8 $\pm{}$ 2 & $54 \pm{}$ 13 & $3_{-2}^{+4}$ & $2_{-1}^{+6}$ & $0.7\%$ & $3.0\%$ &  & 1 & 2\\
NGC 6284 & $<13$ & $<110$ & 8 $\pm{}$ 1 & $53 \pm{}$ 6 & $1_{-1}^{+2}$ & $0.9_{-0.5}^{+2.4}$ & $10.7\%$ & $13.4\%$ &  &  & \\
NGC 6626 & $29_{-14}^{+13}$ & $220_{-140}^{+130}$ & 7 $\pm{}$ 1 & $52 \pm{}$ 5 & $7_{-4}^{+11}$ & $6_{-3}^{+18}$ & $0.6\%$ & $1.5\%$ & 5 & 11 & 14\\
NGC 6093 & $13_{-8}^{+9}$ & $100_{-77}^{+91}$ & 7 $\pm{}$ 1 & $47 \pm{}$ 4 & $2_{-1}^{+3}$ & $1_{-1}^{+4}$ & $2.7\%$ & $5.5\%$ &  & 1 & 3\\
NGC 7089 & $<7$ & $<59$ & 7 $\pm{}$ 1 & $46 \pm{}$ 5 & $1_{-1}^{+2}$ & $1_{-1}^{+3}$ & $45.3\%$ & $30.5\%$ &  &  & 10\\
NGC 5286 & $11_{-7}^{+8}$ & $84_{-63}^{+78}$ & 6 $\pm{}$ 1 & $43 \pm{}$ 4 & $1_{-1}^{+2}$ & $1_{-1}^{+4}$ & $3.9\%$ & $6.4\%$ &  &  & \\
NGC 6752 & $3_{-2}^{+3}$ & $23_{-19}^{+24}$ & 6 $\pm{}$ 2 & $40 \pm{}$ 11 & $10_{-7}^{+18}$ & $8_{-6}^{+25}$ & $54.8\%$ & $39.4\%$ & 3 & 5 & 9\\
NGC 6541 & $8_{-5}^{+7}$ & $68_{-54}^{+59}$ & 6 $\pm{}$ 1 & $40 \pm{}$ 6 & $3_{-2}^{+5}$ & $2_{-1}^{+7}$ & $6.8\%$ & $10.2\%$ &  &  & \\
NGC 6453 & $<34$ & $<290$ & 6 $\pm{}$ 1 & $39 \pm{}$ 8 & $2_{-1}^{+3}$ & $1_{-1}^{+4}$ & $0.4\%$ & $1.9\%$ &  &  & \\
NGC 6522 & $<14$ & $<120$ & 5 $\pm{}$ 1 & $38 \pm{}$ 8 & $3_{-2}^{+6}$ & $2_{-2}^{+8}$ & $6.7\%$ & $7.3\%$ & 4 & 10 & 6\\
NGC 6517 & $<10$ & $<83$ & 5 $\pm{}$ 1 & $37 \pm{}$ 10 & $2_{-1}^{+3}$ & $1_{-1}^{+4}$ & $20.9\%$ & $16.0\%$ & 4 & 10 & 21\\
NGC 7099 & $<5$ & $<38$ & 5 $\pm{}$ 1 & $36 \pm{}$ 8 & $2_{-1}^{+4}$ & $2_{-1}^{+6}$ & $74.5\%$ & $50.4\%$ & 1 & 1 & 2\\
NGC 6139 & $17_{-10}^{+12}$ & $130_{-98}^{+110}$ & 5 $\pm{}$ 1 & $35 \pm{}$ 7 & $1_{-1}^{+3}$ & $1_{-1}^{+4}$ & $1.0\%$ & $2.7\%$ &  &  & \\
NGC 6864 & $<16$ & $<130$ & 5 $\pm{}$ 1 & $35 \pm{}$ 7 & $0.3_{-0.2}^{+0.6}$ & $0.2_{-0.2}^{+0.9}$ & $5.5\%$ & $6.8\%$ &  &  & \\
NGC 6528 & $8_{-6}^{+7}$ & $61_{-48}^{+65}$ & 5 $\pm{}$ 1 & $33 \pm{}$ 7 & $2_{-2}^{+4}$ & $2_{-1}^{+5}$ & $6.4\%$ & $9.3\%$ &  &  & \\
NGC 6341 & $3_{-2}^{+4}$ & $28_{-23}^{+30}$ & 5 $\pm{}$ 1 & $33 \pm{}$ 3 & $2_{-1}^{+4}$ & $1_{-1}^{+5}$ & $38.0\%$ & $25.7\%$ &  & 1 & 2\\
NGC 6273 & $<22$ & $<180$ & 4 $\pm{}$ 1 & $28 \pm{}$ 5 & $2_{-1}^{+3}$ & $1_{-1}^{+4}$ & $1.4\%$ & $3.1\%$ &  &  & \\
NGC 5272 & $<5$ & $<45$ & 4 $\pm{}$ 1 & $28 \pm{}$ 3 & $1_{-1}^{+2}$ & $0.7_{-0.5}^{+3.0}$ & $49.5\%$ & $30.4\%$ & 2 & 4 & 5\\
NGC 5694 & $<28$ & $<240$ & 4 $\pm{}$ 1 & $27 \pm{}$ 4 & $0.09_{-0.06}^{+0.22}$ & $0.07_{-0.05}^{+0.27}$ & $0.9\%$ & $2.1\%$ &  &  & \\
NGC 6256 & $<8$ & $<64$ & 4 $\pm{}$ 1 & $26 \pm{}$ 10 & $2_{-1}^{+5}$ & $1_{-1}^{+6}$ & $24.4\%$ & $16.1\%$ &  &  & \\
NGC 5904 & $4_{-3}^{+4}$ & $30_{-24}^{+34}$ & 4 $\pm{}$ 1 & $25 \pm{}$ 4 & $2_{-1}^{+4}$ & $1_{-1}^{+5}$ & $24.6\%$ & $19.1\%$ & 2 & 5 & 7\\
NGC 6638 & $<21$ & $<190$ & 3 $\pm{}$ 1 & $23 \pm{}$ 4 & $0.9_{-0.7}^{+2.5}$ & $0.7_{-0.5}^{+2.3}$ & $1.3\%$ & $3.0\%$ &  &  & \\
NGC 5946 & $<9$ & $<72$ & 3 $\pm{}$ 1 & $23 \pm{}$ 5 & $0.9_{-0.7}^{+2.1}$ & $0.7_{-0.6}^{+3.0}$ & $16.0\%$ & $12.2\%$ &  &  & \\
\\
\end{tabular}

\end{table*}

\begin{table*}
	\centering
        \ContinuedFloat
	\caption{Continued.}
        \renewcommand{\arraystretch}{1.25} % Default value: 1
    %\label{tab:Table_2}
	\begin{tabular}{llllllllllll}
\hline
GC & $N^{A}_{\gamma}$ & $N^{B}_{\gamma}$ & $N_{\Gamma}^A$ & $N_{\Gamma}^B$ & $F_{\gamma}^{A}$ & $F_{\gamma}^{B}$ & $p_{b}^{A}$ & $p_{b}^{B}$ & $N^{A}_{\rm Radio}$ & $N^{B}_{\rm Radio}$ & $N_{\rm PSR, R}$\\ &  &  &  &  & \multicolumn{2}{c}{($10^{-12}$ erg s$^{-1}$ cm$^{-2}$)} & & & \multicolumn{2}{c}{($> S_{\rm min}$)} &  \\\hline
NGC 6333 & $<8$ & $<72$ & 3 $\pm{}$ 1 & $23 \pm{}$ 6 & $1_{-1}^{+3}$ & $0.9_{-0.7}^{+4.2}$ & $18.2\%$ & $12.8\%$ &  &  & \\
NGC 6402 & $10_{-6}^{+8}$ & $82_{-62}^{+71}$ & 3 $\pm{}$ 1 & $22 \pm{}$ 4 & $0.9_{-0.6}^{+2.4}$ & $0.7_{-0.5}^{+2.8}$ & $2.5\%$ & $3.6\%$ &  &  & 5\\
NGC 6304 & $6_{-4}^{+6}$ & $47_{-39}^{+53}$ & 3 $\pm{}$ 1 & $22 \pm{}$ 5 & $2_{-2}^{+5}$ & $2_{-1}^{+8}$ & $8.6\%$ & $7.9\%$ &  &  & \\
NGC 6325 & $<6$ & $<57$ & 3 $\pm{}$ 1 & $21 \pm{}$ 6 & $1_{-1}^{+4}$ & $1_{-1}^{+6}$ & $32.0\%$ & $20.8\%$ &  &  & \\
NGC 6380 & $<9$ & $<72$ & 3 $\pm{}$ 1 & $21 \pm{}$ 7 & $0.8_{-0.7}^{+2.1}$ & $0.6_{-0.5}^{+2.7}$ & $15.1\%$ & $11.1\%$ &  &  & \\
NGC 1904 & $16_{-9}^{+11}$ & $130_{-89}^{+99}$ & 3 $\pm{}$ 1 & $21 \pm{}$ 2 & $0.4_{-0.3}^{+1.0}$ & $0.3_{-0.2}^{+1.3}$ & $0.6\%$ & $1.7\%$ &  &  & 1\\
NGC 6544 & $3_{-2}^{+3}$ & $19_{-16}^{+28}$ & 3 $\pm{}$ 1 & $21 \pm{}$ 7 & $12_{-10}^{+30}$ & $8_{-7}^{+44}$ & $43.6\%$ & $26.0\%$ &  &  & 3\\
NGC 6558 & $<6$ & $<47$ & 3 $\pm{}$ 1 & $20 \pm{}$ 3 & $1_{-1}^{+4}$ & $0.9_{-0.7}^{+4.1}$ & $35.8\%$ & $24.4\%$ &  &  & \\
NGC 6355 & $<9$ & $<78$ & 3 $\pm{}$ 1 & $19 \pm{}$ 5 & $0.9_{-0.7}^{+2.6}$ & $0.6_{-0.5}^{+2.8}$ & $13.1\%$ & $9.0\%$ &  &  & \\
NGC 6642 & $<8$ & $<67$ & 3 $\pm{}$ 1 & $19 \pm{}$ 4 & $1.0_{-0.8}^{+2.8}$ & $0.8_{-0.6}^{+3.6}$ & $15.2\%$ & $13.5\%$ &  &  & \\
NGC 5139 & $9_{-6}^{+6}$ & $75_{-57}^{+60}$ & 3 $\pm{}$ 1 & $19 \pm{}$ 3 & $2_{-2}^{+6}$ & $2_{-1}^{+8}$ & $1.6\%$ & $3.9\%$ & 2 & 3 & 19\\
NGC 6637 & $<17$ & $<150$ & 3 $\pm{}$ 1 & $19 \pm{}$ 4 & $0.9_{-0.7}^{+2.1}$ & $0.6_{-0.4}^{+2.9}$ & $2.1\%$ & $2.2\%$ &  &  & 2\\
NGC 6356 & $<19$ & $<160$ & 3 $\pm{}$ 1 & $18 \pm{}$ 2 & $0.3_{-0.2}^{+0.7}$ & $0.2_{-0.1}^{+0.8}$ & $1.8\%$ & $2.8\%$ &  &  & \\
NGC 6397 & $2_{-1}^{+1}$ & $10_{-8}^{+15}$ & 3 $\pm{}$ 1 & $18 \pm{}$ 3 & $10_{-7}^{+23}$ & $7_{-5}^{+30}$ & $76.3\%$ & $51.7\%$ & 2 & 3 & 2\\
NGC 6656 & $3_{-2}^{+3}$ & $22_{-19}^{+29}$ & 2 $\pm{}$ 1 & $17 \pm{}$ 4 & $5_{-4}^{+19}$ & $4_{-3}^{+17}$ & $25.6\%$ & $19.4\%$ & 2 & 3 & 4\\
NGC 6316 & $41_{-18}^{+18}$ & $320_{-200}^{+180}$ & 2 $\pm{}$ 1 & $17 \pm{}$ 3 & $0.5_{-0.4}^{+1.4}$ & $0.4_{-0.3}^{+1.6}$ & $<0.1\%$ & $0.5\%$ &  &  & 1\\
NGC 6553 & $<5$ & $<39$ & 2 $\pm{}$ 1 & $16 \pm{}$ 4 & $2_{-2}^{+6}$ & $1_{-1}^{+6}$ & $41.4\%$ & $25.3\%$ &  &  & \\
NGC 6205 & $2_{-1}^{+2}$ & $14_{-12}^{+18}$ & 2 $\pm{}$ 1 & $16 \pm{}$ 3 & $1.0_{-0.7}^{+2.7}$ & $0.7_{-0.5}^{+3.6}$ & $56.9\%$ & $32.4\%$ & 2 & 2 & 9\\
NGC 5927 & $<7$ & $<56$ & 2 $\pm{}$ 1 & $16 \pm{}$ 2 & $0.8_{-0.6}^{+2.6}$ & $0.6_{-0.5}^{+3.0}$ & $22.0\%$ & $13.8\%$ &  &  & \\
NGC 5986 & $<8$ & $<66$ & 2 $\pm{}$ 1 & $15 \pm{}$ 2 & $0.4_{-0.3}^{+1.5}$ & $0.3_{-0.3}^{+1.5}$ & $13.8\%$ & $8.5\%$ & 1 & 1 & 1\\
Pal 10 & $<10$ & $<82$ & 2 $\pm{}$ 1 & $15 \pm{}$ 7 & $0.6_{-0.5}^{+2.7}$ & $0.4_{-0.4}^{+2.7}$ & $9.6\%$ & $9.4\%$ &  &  & \\
NGC 6760 & $<8$ & $<71$ & 2 $\pm{}$ 1 & $15 \pm{}$ 4 & $0.7_{-0.6}^{+2.2}$ & $0.5_{-0.4}^{+2.3}$ & $12.2\%$ & $10.8\%$ &  &  & 2\\
NGC 6569 & $<31$ & $<270$ & 2 $\pm{}$ 1 & $14 \pm{}$ 5 & $0.4_{-0.3}^{+1.5}$ & $0.3_{-0.3}^{+1.8}$ & $0.2\%$ & $1.4\%$ &  &  & \\
NGC 6229 & $<24$ & $<200$ & 2 $\pm{}$ 1 & $13 \pm{}$ 4 & $0.05_{-0.04}^{+0.17}$ & $0.03_{-0.03}^{+0.19}$ & $0.6\%$ & $1.3\%$ &  &  & \\
NGC 6342 & $<22$ & $<180$ & 2 $\pm{}$ 1 & $13 \pm{}$ 3 & $0.7_{-0.6}^{+2.1}$ & $0.5_{-0.4}^{+2.6}$ & $0.8\%$ & $0.8\%$ &  &  & 2\\
NGC 6401 & $18_{-10}^{+11}$ & $140_{-110}^{+110}$ & 2 $\pm{}$ 1 & $13 \pm{}$ 2 & $0.6_{-0.5}^{+2.3}$ & $0.5_{-0.4}^{+2.8}$ & $0.1\%$ & $0.9\%$ &  &  & \\
NGC 6539 & $<33$ & $<280$ & 2 $\pm{}$ 1 & $12 \pm{}$ 5 & $0.6_{-0.5}^{+2.5}$ & $0.4_{-0.3}^{+2.4}$ & $<0.1\%$ & $0.3\%$ &  &  & 1\\
NGC 6717 & $6_{-4}^{+6}$ & $52_{-41}^{+49}$ & 2 $\pm{}$ 1 & $12 \pm{}$ 4 & $0.6_{-0.6}^{+2.3}$ & $0.4_{-0.4}^{+3.0}$ & $4.3\%$ & $2.9\%$ &  &  & \\
NGC 6287 & $<6$ & $<52$ & 2 $\pm{}$ 1 & $12 \pm{}$ 2 & $0.6_{-0.5}^{+2.2}$ & $0.4_{-0.3}^{+2.7}$ & $20.5\%$ & $10.4\%$ &  &  & \\
NGC 5024 & $<10$ & $<84$ & 2 $\pm{}$ 1 & $11 \pm{}$ 2 & $0.1_{-0.08}^{+0.39}$ & $0.07_{-0.06}^{+0.51}$ & $5.8\%$ & $4.5\%$ &  &  & 5\\
NGC 6254 & $<7$ & $<61$ & 2 $\pm{}$ 1 & $11 \pm{}$ 1 & $1_{-1}^{+6}$ & $0.9_{-0.7}^{+5.6}$ & $13.9\%$ & $9.2\%$ & 1 & 1 & 2\\
NGC 6712 & $7_{-5}^{+7}$ & $59_{-47}^{+58}$ & 2 $\pm{}$ 1 & $11 \pm{}$ 1 & $0.6_{-0.5}^{+2.6}$ & $0.4_{-0.3}^{+2.7}$ & $3.5\%$ & $3.2\%$ & 1 & 1 & 1\\
NGC 6934 & $<11$ & $<96$ & 1 $\pm{}$ 1 & $10 \pm{}$ 2 & $0.1_{-0.1}^{+0.6}$ & $0.09_{-0.08}^{+0.73}$ & $3.6\%$ & $3.8\%$ &  &  & \\
NGC 6779 & $<8$ & $<65$ & 1 $\pm{}$ 1 & $10 \pm{}$ 3 & $0.3_{-0.2}^{+1.2}$ & $0.2_{-0.2}^{+1.4}$ & $11.3\%$ & $6.1\%$ &  &  & \\
NGC 6121 & $<2$ & $<11$ & 1 $\pm{}$ 1 & $10 \pm{}$ 3 & $8_{-7}^{+41}$ & $6_{-5}^{+40}$ & $92.2\%$ & $71.7\%$ &  &  & 1\\
Ter 2 & $15_{-9}^{+13}$ & $120_{-87}^{+110}$ & 1 $\pm{}$ 1 & $9 \pm{}$ 6 & $0.5_{-0.4}^{+2.5}$ & $0.3_{-0.3}^{+2.1}$ & $0.2\%$ & $1.2\%$ &  &  & \\
Pal 11 & $<13$ & $<120$ & 1 $\pm{}$ 1 & $9 \pm{}$ 3 & $0.1_{-0.1}^{+0.6}$ & $0.09_{-0.07}^{+0.57}$ & $1.1\%$ & $1.7\%$ &  &  & \\
NGC 5634 & $<19$ & $<170$ & 1 $\pm{}$ 1 & $8 \pm{}$ 3 & $0.03_{-0.03}^{+0.17}$ & $0.02_{-0.02}^{+0.2}$ & $1.0\%$ & $0.8\%$ &  &  & \\
NGC 4147 & $<92$ & $<760$ & 1 $\pm{}$ 1 & $8 \pm{}$ 3 & $0.06_{-0.06}^{+0.45}$ & $0.04_{-0.04}^{+0.31}$ & $<0.1\%$ & $<0.1\%$ &  &  & \\
Pal 6 & $7_{-6}^{+8}$ & $58_{-50}^{+73}$ & 1 $\pm{}$ 1 & $7 \pm{}$ 3 & $0.4_{-0.4}^{+2.5}$ & $0.3_{-0.2}^{+2.4}$ & $2.4\%$ & $2.6\%$ &  &  & \\
\\
\end{tabular}

\end{table*}

\begin{table*}
	\centering
        \ContinuedFloat
	\caption{Continued.}
        \renewcommand{\arraystretch}{1.25} % Default value: 1
    %\label{tab:Table_2}
	\begin{tabular}{llllllllllll}
\hline
GC & $N^{A}_{\gamma}$ & $N^{B}_{\gamma}$ & $N_{\Gamma}^A$ & $N_{\Gamma}^B$ & $F_{\gamma}^{A}$ & $F_{\gamma}^{B}$ & $p_{b}^{A}$ & $p_{b}^{B}$ & $N^{A}_{\rm Radio}$ & $N^{B}_{\rm Radio}$ & $N_{\rm PSR, R}$\\ &  &  &  &  & \multicolumn{2}{c}{($10^{-12}$ erg s$^{-1}$ cm$^{-2}$)} & & & \multicolumn{2}{c}{($> S_{\rm min}$)} &  \\\hline
NGC 1261 & $<9$ & $<73$ & 1 $\pm{}$ 1 & $7 \pm{}$ 2 & $0.07_{-0.06}^{+0.37}$ & $0.04_{-0.04}^{+0.35}$ & $6.3\%$ & $2.5\%$ &  &  & \\
NGC 6218 & $2_{-1}^{+2}$ & $16_{-13}^{+19}$ & < 1 & $7 \pm{}$ 2 & $< 5$ & $0.4_{-0.4}^{+3.4}$ & $28.0\%$ & $13.1\%$ &  &  & 2\\
NGC 6584 & $<8$ & $<63$ & < 1 & $6 \pm{}$ 2 & $< 0.6$ & $0.06_{-0.05}^{+0.55}$ & $7.2\%$ & $3.9\%$ &  &  & \\
NGC 6723 & $<6$ & $<49$ & < 1 & $6 \pm{}$ 2 & $< 2$ & $0.2_{-0.1}^{+1.2}$ & $11.7\%$ & $8.2\%$ &  &  & \\
IC 1276 & $<4$ & $<31$ & < 1 & $5 \pm{}$ 3 & $< 6$ & $0.4_{-0.4}^{+4.3}$ & $35.6\%$ & $15.3\%$ &  &  & \\
NGC 3201 & $<3$ & $<20$ & < 1 & $5 \pm{}$ 1 & $< 5$ & $0.3_{-0.3}^{+3.4}$ & $57.1\%$ & $21.7\%$ &  &  & \\
NGC 6171 & $<4$ & $<36$ & < 1 & $5 \pm{}$ 1 & $< 4$ & $0.2_{-0.2}^{+2.1}$ & $27.3\%$ & $8.2\%$ &  &  & \\
NGC 6352 & $<4$ & $<30$ & < 1 & $5 \pm{}$ 1 & $< 4$ & $0.2_{-0.2}^{+2.2}$ & $38.0\%$ & $13.3\%$ &  &  & \\
NGC 4590 & $<12$ & $<110$ & < 1 & $4 \pm{}$ 1 & $< 0.9$ & $0.05_{-0.05}^{+0.68}$ & $1.8\%$ & $1.1\%$ &  &  & \\
NGC 6235 & $<9$ & $<80$ & < 1 & $4 \pm{}$ 1 & $< 0.8$ & $0.04_{-0.04}^{+0.41}$ & $4.6\%$ & $1.9\%$ &  &  & \\
NGC 6366 & $<3$ & $<24$ & < 1 & $4 \pm{}$ 1 & $< 10$ & $0.4_{-0.4}^{+4.7}$ & $49.2\%$ & $18.3\%$ &  &  & \\
NGC 6981 & $<12$ & $<110$ & < 1 & $4 \pm{}$ 1 & $< 0.4$ & $0.02_{-0.02}^{+0.2}$ & $1.6\%$ & $0.9\%$ &  &  & \\
NGC 6362 & $<5$ & $<40$ & < 1 & $4 \pm{}$ 1 & $< 2$ & $0.08_{-0.07}^{+1.11}$ & $19.5\%$ & $8.0\%$ &  &  & \\
NGC 2298 & $<14$ & $<130$ & < 1 & $4 \pm{}$ 1 & $< 1$ & $0.05_{-0.04}^{+0.71}$ & $1.2\%$ & $0.6\%$ &  &  & \\
Ton 2 & $<4$ & $<32$ & < 1 & $4 \pm{}$ 2 & $< 2$ & $0.09_{-0.09}^{+1.27}$ & $29.6\%$ & $9.0\%$ &  &  & \\
Pal 8 & $<9$ & $<79$ & < 1 & $4 \pm{}$ 1 & $< 0.8$ & $0.03_{-0.03}^{+0.41}$ & $4.5\%$ & $1.8\%$ &  &  & \\
NGC 6809 & $<3$ & $<24$ & < 1 & $3 \pm{}$ 1 & $< 4$ & $0.1_{-0.1}^{+2.2}$ & $48.3\%$ & $12.7\%$ &  &  & \\
NGC 6144 & $<7$ & $<62$ & < 1 & $3 \pm{}$ 1 & $< 2$ & $0.05_{-0.05}^{+0.93}$ & $6.1\%$ & $2.1\%$ &  &  & \\
Ter 9 & $<17$ & $<150$ & < 1 & $2 \pm{}$ 1 & $< 4$ & $0.06_{-0.06}^{+1.18}$ & $1.4\%$ & $0.5\%$ &  &  & \\
Ter 7 & $<26$ & $<230$ & < 1 & $2 \pm{}$ 1 & $< 0.2$ & $0.003_{-0.003}^{+0.063}$ & $0.1\%$ & $0.1\%$ &  &  & \\
NGC 6426 & $<33$ & $<290$ & < 1 & $2 \pm{}$ 1 & $< 0.3$ & $0.004_{-0.004}^{+0.089}$ & $<0.1\%$ & $0.3\%$ &  &  & \\
NGC 6101 & $<37$ & $<320$ & < 1 & $2 \pm{}$ 1 & $< 0.5$ & $0.005_{-0.005}^{+0.114}$ & $<0.1\%$ & $<0.1\%$ &  &  & \\
Pal 1 & $<7$ & $<58$ & < 1 & $2 \pm{}$ 1 & $< 1$ & $0.008_{-0.008}^{+0.182}$ & $8.0\%$ & $1.3\%$ &  &  & \\
NGC 5897 & $<9$ & $<76$ & < 1 & $2 \pm{}$ 1 & $< 0.7$ & $0.006_{-0.006}^{+0.189}$ & $5.0\%$ & $0.8\%$ &  &  & \\
IC 4499 & $<15$ & $<130$ & < 1 & $2 \pm{}$ 1 & $< 0.3$ & $0.002_{-0.002}^{+0.078}$ & $1.2\%$ & $0.5\%$ &  &  & \\
NGC 288 & $<4$ & $<33$ & < 1 & $2 \pm{}$ 1 & $< 1$ & $0.009_{-0.008}^{+0.268}$ & $29.2\%$ & $4.1\%$ &  &  & \\
HP 1 & $<35$ & $<290$ & < 1 & $1 \pm{}$ 1 & $< 2$ & $0.01_{-0.01}^{+0.37}$ & $0.1\%$ & $0.2\%$ &  &  & \\
NGC 6496 & $<7$ & $<60$ & < 1 & $1 \pm{}$ 1 & $< 1$ & $0.009_{-0.009}^{+0.312}$ & $8.2\%$ & $1.8\%$ &  &  & \\
Pal 12 & $<15$ & $<130$ & < 1 & $1 \pm{}$ 1 & $< 0.4$ & $0.002_{-0.002}^{+0.05}$ & $0.7\%$ & $0.2\%$ &  &  & \\
NGC 6535 & $<5$ & $<38$ & < 1 & $1 \pm{}$ 1 & $< 3$ & $0.01_{-0.01}^{+0.42}$ & $22.6\%$ & $1.7\%$ &  &  & \\
Ter 1 & $9_{-6}^{+6}$ & $71_{-54}^{+62}$ & < 1 & < 1 & $< 4$ & $< 0.5$ & $0.7\%$ & $0.1\%$ &  &  & 8\\
NGC 5466 & $<9$ & $<72$ & < 1 & < 1 & $< 0.5$ & $< 0.05$ & $5.1\%$ & $0.5\%$ &  &  & \\
NGC 4372 & $<4$ & $<38$ & < 1 & < 1 & $< 4$ & $< 0.4$ & $23.0\%$ & $2.1\%$ &  &  & \\
NGC 7492 & $<15$ & $<130$ & < 1 & < 1 & $< 0.2$ & $< 0.02$ & $1.2\%$ & $0.6\%$ &  &  & \\
NGC 5053 & $<9$ & $<79$ & < 1 & < 1 & $< 0.4$ & $< 0.04$ & $3.8\%$ & $0.3\%$ &  &  & \\
Arp 2 & $<29$ & $<240$ & < 1 & < 1 & $< 0.2$ & $< 0.02$ & $<0.1\%$ & $<0.1\%$ &  &  & \\
Pal 5 & $<100$ & $<850$ & < 1 & < 1 & $< 0.2$ & $< 0.03$ & $<0.1\%$ & $<0.1\%$ &  &  & \\
Pal 13 & $<22$ & $<190$ & < 1 & < 1 & $< 0.2$ & $< 0.03$ & $0.3\%$ & $0.1\%$ &  &  & \\
\\
\end{tabular}

\end{table*}

\end{document}